\documentclass[acmlarge,screen,authorversion,nonacm,review=false,timestamp=false]{acm/acmart}

\usepackage{amsmath,amsfonts}
\usepackage{algorithmic}
\usepackage{graphicx}
\usepackage{longtable}
\usepackage{multirow}
\usepackage{textcomp}
\usepackage{hyperref}

\newcommand{\revision}[1]{{#1}}

\AtBeginDocument{%
  }

\usepackage{fancyhdr}

\fancypagestyle{firststyle}
{
   \fancyhf{}
   \fancyhead[C]{\textit{\textbf{Preprint --- Under review in the ACM Transactions on Computing for Healthcare (HEALTH) journal.}}}
}
\setcopyright{acmlicensed}
\copyrightyear{2024}
\acmYear{2024}
\acmDOI{XXXXXXX.XXXXXXX}

\acmJournal{HEALTH}
\begin{document}

\title{Health Misinformation in Social Networks: A Survey of IT Approaches}

\author{Vasiliki Papanikou}
\affiliation{
  \institution{Computer Science and Engineering Department (CSE), University of Ioannina (UOI)}
  \country{Greece}
}
\email{v.papanikou@uoi.gr}

\author{Panagiotis Papadakos}
\orcid{0000-0001-8926-4229}
\email{papadako@ics.forth.gr}
\affiliation{
  \institution{Computer Science and Engineering Department (CSE), University of Ioannina (UOI), \& Institute of Computer Science (ICS), Foundation for Research and Technology - Hellas (FORTH)}
  \country{Greece}
}

%\affiliation{%
%  \institution{Institute of Computer Science of the Foundation for Research and Technology - Hellas, Crete}
%  \streetaddress{P.O. Box GR-70013}
%  \city{Ioannina}
%  \country{Greece}

\author{Theodora Karamanidou}
\email{Theodora.Karamanidou@pfizer.com}
\orcid{0000-0002-9766-983X}
\author{Thanos G. Stavropoulos}
\email{Thanos.Stavropoulos@pfizer.com}
\orcid{0000-0003-2389-4329}
\affiliation{
 \institution{Center for Digital Innovation (CDI) of Pfizer}
 \country{Greece}}

\author{Evaggelia Pitoura}
\email{pitoura@uoi.gr}
\orcid{0000-0002-3775-4995}
\author{Panayiotis Tsaparas}
\email{tsap@uoi.gr}
\orcid{0000-0002-3490-1507}
\affiliation{%
  \institution{Computer Science and Engineering Department (CSE), University of Ioannina (UOI)}
  \country{Greece}
}

\renewcommand{\shortauthors}{Papanikou et al.}

\begin{abstract}
    In this paper, we present a comprehensive survey on the pervasive issue of medical misinformation in social networks from the perspective of information technology.  The survey aims at providing a systematic review of related research and helping researchers and practitioners navigate through this fast-changing field.  Specifically, we first present manual and automatic approaches for fact-checking. We then explore fake news detection methods, using content, propagation features, or source features, as well as mitigation approaches for countering the spread of misinformation. We also provide a detailed list of several datasets on health misinformation and of publicly available tools. We conclude the survey with a discussion on the open challenges and future research directions in the battle against health misinformation. 
\end{abstract}

\begin{CCSXML}
<ccs2012>
   <concept>
       <concept_id>10010147.10010257</concept_id>
       <concept_desc>Computing methodologies~Machine learning</concept_desc>
       <concept_significance>500</concept_significance>
       </concept>
   <concept>
       <concept_id>10010147.10010178.10010187</concept_id>
       <concept_desc>Computing methodologies~Knowledge representation and reasoning</concept_desc>
       <concept_significance>500</concept_significance>
       </concept>
   <concept>
       <concept_id>10002951.10003260</concept_id>
       <concept_desc>Information systems~World Wide Web</concept_desc>
       <concept_significance>500</concept_significance>
       </concept>
   <concept>
       <concept_id>10002951.10003260.10003261</concept_id>
       <concept_desc>Information systems~Web searching and information discovery</concept_desc>
       <concept_significance>500</concept_significance>
       </concept>
   <concept>
       <concept_id>10002951.10003260.10003282</concept_id>
       <concept_desc>Information systems~Web applications</concept_desc>
       <concept_significance>500</concept_significance>
       </concept>
   <concept>
       <concept_id>10002944.10011122.10002945</concept_id>
       <concept_desc>General and reference~Surveys and overviews</concept_desc>
       <concept_significance>500</concept_significance>
       </concept>
 </ccs2012>
\end{CCSXML}

\ccsdesc[500]{Computing methodologies~Machine learning}
\ccsdesc[500]{Computing methodologies~Knowledge representation and reasoning}
\ccsdesc[500]{Information systems~World Wide Web}
\ccsdesc[500]{Information systems~Web searching and information discovery}
\ccsdesc[500]{Information systems~Web applications}
\ccsdesc[500]{General and reference~Surveys and overviews}

\keywords{misinformation, fake news, fact-checking, medical domain, COVID-19, information technology approaches, content-based methods, propagation models, source credibility, survey}

%\received{...}
%\received[revised]{...}
%\received[accepted]{...}

\maketitle

\thispagestyle{firststyle} 

\section{Introduction}
\label{sec:introduction}
The spread of misinformation online, most commonly known as fake news,
%especially regarding healthcare and health issues,
is an important issue that has become more pronounced in the last two decades due to the prevalence of social media. Platforms like Twitter, Reddit, and Facebook, have been commonly identified as the main channels for propagating misinformation and have been criticized for not acting on addressing the conditions that permit the circulation and amplification of false information \cite{cotter2022fact}. Such misinformation includes false claims and non fact-checked news items, that originate from sources of questionable credibility \cite{molina2021fake}.
%During the recent pandemic, the extent of misinformation online led the World Health Organization (WHO) to declare an ``infodemic", meaning an overabundance of information – some accurate and some not – that makes it hard for people to find trustworthy sources and reliable guidance when they need it\footnote{\url{https://www.who.int/health-topics/infodemic}}.

The problem of misinformation becomes critical when it pertains to healthcare and health issues, since it puts lives and the public health at risk.
One of the first cases of widely spread misinformation in the medical domain is the falsehood that the MMR vaccine (Measles, Mumps, Rubella) causes autism \cite{mesquita2020infodemia}. The falsehood originated from a fraudulent article titled “Ileal-lymphoid-nodular hyperplasia, non-specific colitis, and pervasive developmental disorder in children” published in the prestigious Lancet journal in 1998 \cite{WHO22, sundstrom2021correcting}. This study turned tens of thousands of parents against the vaccine, and as a result, in 2020, many countries, including the United Kingdom, Greece, Venezuela, and Brazil, lost their measles elimination status. In 2020, twenty-two years after publishing this study Lancet retracted the paper \cite{wuhrl2021claim}.
%Other
Examples of medical fake news that have spread on social media include the oncogenic effects of antihypertensive drugs, which caused several patients to stop using them,
%Also, a large collection of tweets had propagated
and misinformation about the Human Papillomavirus (HPV) vaccines, which resulted in half of the population in South Carolina not having completed the vaccination series \cite{sundstrom2021correcting, weinzierl2022vaccinelies}.

\revision{
The issue of fake news was exacerbated during the recent COVID-19 pandemic,
where it became clear that the enemy %in the COVID-19 pandemic
was not only the virus but also the abundance of misinformation leaked on social media and the web, even from prominent public figures, endangering human lives\footnote{\url{https://www.europol.europa.eu/covid-19/covid-19-fake-news}}.
The extent of online health misinformation led the World Health Organization (WHO) to declare an ``infodemic", that is, ``an overabundance of information – some accurate and some not – that makes it hard for people to find trustworthy sources and reliable guidance when they need it''\footnote{\url{https://www.who.int/health-topics/infodemic}}.
%The misinformation propagation escalated recently after the outbreak of the COVID-19 virus. During the COVID-19 pandemic,
%During the pandemic, it became clear that the enemy %in the COVID-19 pandemic was not only the virus but also the abundance of misinformation leaked on social media and the web, even from prominent public figures, endangering human lives\footnote{\url{https://www.europol.europa.eu/covid-19/covid-19-fake-news}}.
%It was an emergency to deal with this infodemic since the inaccurate information that was spreading on social media
%The repercussions of widespread health misinformation on social media are severe.
The repercussions of the infodemic are severe: It results in ``an increase in erroneous interpretation of scientific knowledge, opinion polarization, escalating fear, and panic or decreased access to health care”\cite{WHO22}, and it is detrimental to public health, impacting healthcare utilization and cost, and medical non-compliance \cite{treharne2020defining}.
Indicatively, in a recent poll\footnote{\url{https://debeaumont.org/wp-content/uploads/2023/03/misinformation-poll-brief-FINAL.pdf}} among 806 physicians, three out of four of the physicians said that medical misinformation has hindered their ability to treat COVID-19 patients, and 44\% of them estimated that more than half of COVID-19 information they receive from patients is misinformation. Furthermore, more than two-thirds of the physicians mentioned that the problem of misinformation extends beyond COVID, to areas such as weight loss, dietary supplements, mental health, and other vaccines.
}

%The repercussions of health misinformation on social media are severe. They result in ``an increase in erroneous interpretation of scientific knowledge, opinion polarization, escalating fear, and panic or decreased access to health care”\cite{WHO22}, and they are detrimental to public health, impacting healthcare utilization and cost, and medical non-compliance \cite{treharne2020defining}. In a recent poll\footnote{\url{https://debeaumont.org/wp-content/uploads/2023/03/misinformation-poll-brief-FINAL.pdf}} among 806 physicians and 2,210 randomly selected adults, three out of four of the physicians said that medical misinformation has hindered their ability to treat COVID-19 patients, and 44\% of them estimate that more than half of COVID-19 information they receive from patients is misinformation. Furthermore, more than two-thirds of the physicians mentioned that misinformation is also a problem in the areas of weight loss, dietary supplements, mental health, and other vaccines. In the same poll, almost half of the physicians rank scientific papers as the most trusted source of information for medical care and treatment, while half of the non-expert group trusts Internet searches. However, both groups show limited trust in social media, with 0\% of the physicians choosing social media as the most trusted source.

As a response to this wave of dangerous misinformation, there was a call-to-arms to construct tools for weathering the infodemic. There was a strong effort from the scientific community to construct and make publicly available datasets with
%
%As a response to this wave of dangerous misinformation, the  scientific community constructed datasets and tools to index and retrieve
valid scholarly information relevant to COVID-19 and coronaviruses in general.
A well-known dataset is  the COVID-19 Open Research Dataset (CORD-19) \cite{wang2020cord} and the Covidex search engine \cite{zhang2020rapidly} built on top of it.
In addition, most fact-checking organizations had to grow and dedicate more of their resources to COVID-19 misinformation \cite{ginsberg2021report}. As a result, several fact-checking services were developed targeting the COVID-19 case, aiming at assessing the validity of claims and the credibility of sources.
However, it is becoming increasingly challenging and expensive to identify fake news and claims through manual inspection, due to the speed at which information (both valid and non-valid) spreads on the media, and especially on social media. In a WHO survey on COVID-19 fact-checkers\footnote{WHO cooperates with a global network of 200 fact-checking organizations \url{https://covid19misinfo.org/fact-checking/covid-19-fact-checkers/}
}, most fact-checking organizations reported that their leading challenge was keeping up with rapidly changing science and fact-checking health misinformation \cite{purnat2021}.
%This highlights the importance of automatic, data-driven approaches that can complement or supplant manual efforts.

\revision{The inability of manual approaches to deal with the scale of the problem highlights the importance of automatic, data-driven approaches that can complement or supplant manual efforts. The research efforts of the Information Technology scientific community for addressing misinformation focus on a variety of issues, such as automating claim detection and validation, detecting misinformation using the content and the propagation patterns, identifying users (malicious or not) that instigate or facilitate misinformation, as well as, mitigating misinformation. Most of the approaches develop and apply state-of-the-art Data Science and Machine Learning techniques, trained on large amounts of data. }

%\noindent \textbf{Scope of the survey:}
The research on misinformation is extremely broad, touching several scientific fields. Previous technical surveys on fact-checking and fake news provide a general overview of the current landscape but do not target the medical domain \cite{guo2022survey, zhou2020survey, zeng2021automated}. The few surveys that explicitly focus on misinformation in the medical domain \cite{borges2022infodemics, chen2022combating, suarez2021prevalence, wang2019systematic} provide a high-level approach that is relevant to a broad spectrum of researchers, stakeholders, and decision-makers. This survey complements the previous works by focusing on %fact-checking and fake news detection in social networks for
the medical domain,  and exploring how the different fact-checking and fake news detection techniques have been adapted to this domain from a computer engineering perspective. To facilitate future research on the area it thoroughly describes publicly available domain-specific datasets, paying special attention to the COVID-19 case.

%\noindent \textbf{Methodology:}
We followed an exploratory (snowball) methodology for
%this survey, by collecting and studying several
collecting papers on fact-checking and fake news detection on medical issues. The papers were gathered through Google Scholar by submitting appropriate queries, such as "\textit{medical fake news detection}", "\textit{health misinformation detection on social media}", “\textit{detection of COVID-19 vaccine misinformation}”, “\textit{medical fact check}”, etc. We also used authoritative general surveys on fake news and fact-checking \cite{guo2022survey, zhou2020survey, zeng2021automated} as a starting point for exploring the literature and identifying medical-related approaches. Finally, we explored the citations of the most cited papers, in order to cover the most recent advancements in the area. In total, this survey references more than 200 papers, and describes 24 publicly available datasets
%originating from social media content
and 11 fact-checking tools.

%\noindent \textbf{Roadmap:}
The rest of the survey is structured as follows. Section \ref{sec:overview} provides an overview of the domain and the corresponding definitions, while Section \ref{sec:fc} discusses manual and automatic approaches to fact-checking. The following sections discuss three types of automatic approaches for fake news detection in social media. The first one is based on the content of posts and comments (Section \ref{sec:content}), the second one on the propagation of posts in social networks (Section \ref{sec:prop}), and the third one on the credibility of news sources (Section \ref{sec:sources}). Section \ref{sec:DET} reports fake/valid news datasets originating from social platforms like Twitter and Reddit, and some well-known fact-checking tools and services. Section \ref{sec:mitigation} provides a brief summary of available mitigation approaches. Finally, Section \ref{sec:discuss} provides a discussion and explores future directions, while Section \ref{sec:conc} concludes the survey.

\section{Overview}
\label{sec:overview}

%What is fake news? (different definitions)

There is no universally accepted definition of fake news. An understanding of fake news is attained by considering the dimensions of authenticity, intention, and
news content \cite{Tandoc17,zhou2020survey}. The \textit{authenticity dimension} refers to whether the factual claims in the news item are valid. The \textit{intention  dimension} is about whether there is intention to deceive in creating, or propagating the news item. Finally, the \textit{news dimension} refers to whether the content of the item is about news or not. In case of social media, the news dimension is often not clear,  since blogs and social platforms have allowed non-journalists to reach large audiences, challenging the traditional definition of what news is \cite{Tandoc17}.

There are many information disorders along the dimensions
of authenticity and intention that overlap with fake news.
The term \textit{misinformation} is used to characterize both intentionally and unintentionally false or misleading information, while the term \textit{disinformation} is used for false information that is purposely spread to deceive people \cite{Lazer18}. In this paper, we will consider mainly misinformation.
Malinformation is the deliberate dissemination of true information (e.g., leaking private information) and it is not considered fake news \cite{rastogi2022review}.

There is also a long list of other concepts in the authenticity, intention and news spectrum related to fake news \cite{Tandoc17,zhou2020survey,rastogi2022review}.
Two such concepts are \textit{satire} and \textit{parody} that
use humor or exaggeration to draw attention and often entertain. Both share the assumption that the users are aware that the presentation, or the content is intentionally faux.
Fake information may also be part of \textit{advertising}, or
 \textit{public relations} when
marketing or other persuasive messages are inserted into news articles.
A somewhat similar case is \textit{propaganda} that
refers to news items created by political entities to influence public perceptions.
Other concepts include \textit{hoaxes} referring to  half-truths  made for fun, \textit{rumors} referring to ambiguous stories whose truthfulness never gets confirmed, and \textit{clickbaits} referring to misleading headlines for engaging the audience.

%Difference between fake news in general and %fake news in the medical domain

When it comes to social networks, the \textit{life cycle} of fake news  includes the phases of news creation, publishing on the network, and online propagation.
The author of the news item is the person that first created the news item. The author may be a journalist writing a newspaper article, a scientist writing a scientific article, or in some cases, a normal user. The publisher is the person that first posted the item on the social network platform. In some cases, the publisher is also the author of the item.
Propagation means any kind of reaction that leads to the item getting additional exposure. Depending on the media, reactions include re-postings or retweets, commenting and various other actions of endorsement or disapproval, such as likes, upvotes, or downvotes.

In particular, in the medical domain, misinformation is often created and spread by individuals with no scientific affiliation, that assume the role of expert patients, promote individual autonomy, and challenge state actions \cite{wang2019systematic}. By promoting fear and anxiety, and through the horizontal diffusion of conspiracy theories, they are able to erode in an irreversible way the vertical health communication strategies.

Another major issue regarding medical information, and scientific information in general, is the fact that citizens often have a limited understanding of basic scientific facts and more broadly of the scientific process
\cite{Scheufele19}. Moreover, personal beliefs are often inconsistent with the best available science due to inaccurate perceptions, lack of scientific consensus, or adoption of conspiracy theories.

Fake news propagation in social media is also amplified  by the fact
that it is systemic according to \cite{Gelfert18}. Fake news is designed in such a way so as to pass itself as news to the relevant targeted audience and in fact mislead it, by exploiting the systemic features inherent in the channels of social media. Such features include various cognitive biases and heuristics, that lead in increasing the spread of fake news propagation.

% The next two paragraphs with the statistics are too detailed. I would keep only an high-level idea of how often the different types of fake news appear in the health domain and move it after the definition of fake news types
Various pilot studies have been conducted regarding the typology of health misinformation in social media.
The authors of  \cite{Waszak18} examined health-related misinformative posts from various social network platforms written in Polish during the period 2012-2017.
%The study considered the six misinformation types described in \cite{zhou2020survey}.
In the initial screening, satire, parody, and propaganda were not detected, probably because
they mainly apply to political news.  However, 40\% of the most frequently shared links contained text classified as fake news. The most fallacious content concerned vaccines, while, content about cardiovascular diseases was, in general, well-sourced and informative.
Another study of the sources and types of misinformation about COVID-19 highlighted the
prevalence of fake news in social media \cite{Naeem21}.
In this study, the identified common types of misinformation include false claims, conspiracy theories and
pseudo-scientific health therapies, regarding the diagnosis,
treatment, prevention, origin, and spread of the virus.

%The analyzed topics that attracted most public attention were cancer, neoplasm and vaccinations.

%%medical fake news
%In a pilot study considering health misinformation in the Polish language over Facebook, Twitter,  LinkedIn, and Pinterest social media from 2012 to 2017 \cite{Waszak18}, it was found that  Facebook activities accounted for the majority of engagements.
%An application, BuzzSumo, was used to provide data about social media shares of the most popular web pages.
%The analyzed topics that attracted most public attention were cancer, neoplasm and vaccinations.
%The authors  considered the following six different types of misinformation: satire, parody, fabrication, manipulation, propaganda, and advertising.
%Satire, parody, and propaganda were not found in the initial screening of the news, probably because they mainly apply to political news.
%Regarding fake news, 40\% of the most frequently shared links contained text classified as fake news, with the most fallacious content concerning vaccines. On the opposite, content about cardiovascular diseases was, in general, well-sourced and informative.

Finally, in a high-level overview survey on English social media \cite{suarez2021prevalence}, the authors used the PubMed search engine to explore the prevalence of health misinformation and identify the medical topics that are more susceptible to fake information.
They reported high misinformation content in the smoke, drug, and vaccine categories, moderate misinformation about diets and noncommunicable diseases and pandemics, and low misinformation in medical treatments and surgical treatments, since in this case most information is coming from official accounts.

In the following sections, we focus on fact checking, fake news detection and mitigation in social media platforms with emphasis on the medical domain.
\textit {Fact-checking} is the process of verifying the accuracy and truthfulness of information that is presented as news or as factual. Fact-checking involves researching the claims being made, looking for credible sources of information, and comparing the information to other sources to ensure that it is accurate and truthful.
We categorize fact-checking approaches into manual and automatic approaches and present a thorough analysis of the related methods. Fact-checking is important because it helps to prevent the spread of  misinformation, and it allows people to make informed decisions based on accurate information.

In terms of fake news detection, we provide a systematic categorization and thorough description and analysis of fake news detection approaches.  We categorize fake news detection approaches into those using: (a) the content of the news item, (b) information about its propagation in the social platform, and (c) features of the source, that is, of the publisher or the social media users involved in the propagation. For each approach, we describe in depth the used methods, including the commonly used features in case of traditional Machine Learning (ML) approaches and the involved pipelines in the case of  Deep-Learning (DL) approaches.

The three categories of fake news detection approaches are usually fused, since the study of fake news in social media requires the use of both textual and structural information, along with the user context and preferences, the social context, and any spatio-temporal information available \cite{tang2019learning, shu2020fakenewsnet, dou2021user}.
%This is especially important for classic ML classifiers, where the importance of other feature types increases.
For example, according to \cite{di2022health}, the analysis of health-related content that uses a more informal language can benefit from propagation-network and user-profile features, while more formal medical content can benefit from linguistic-stylistic and linguistic-medical features.  Moreover, the analysis of popular content that generates a high volume of social reactions can benefit from linguistic-emotional features \cite{di2022health}.

Fake news mitigation techniques aim mainly at early detection of fake news and at limiting the spread of misinformation. They include approaches applied at the propagation and the source level. Finally, we describe in detail the available datasets and fact-checking tools and services.

\section{Fact-Checking}
\label{sec:fc}

Fact-checking is the process of verifying the factual accuracy of statements/claims. Historically, it has been associated with journalism, being an important process of media companies for verifying published information related to different types of claims (political, religious, social, etc.).
%claims, religious intolerance, conspiracy theories, etc.
The process of fact-checking is usually done either internally, using resources of the media company or through an external third-party \cite{cotter2022fact}. In a similar manner, there is a verification and validation process of academic work in the scientific community, where the critical study of the prior literature, the soundness of the proposed approaches and methodologies, and the reproducibility and verifiability of the results, are integral to academic research.

During the 2016 US elections, the online media platforms were found susceptible to disseminating disinformation and misinformation \cite{allcott2017social}. This made fact-checking a hot topic across the scientific community, journalists, and online social users. This trend was further amplified due to the pandemic misinformation. As a result, the number and size of fact-checking organizations have grown across the globe and social platforms have started to develop partnerships with them \cite{cotter2022fact}. In addition, the need for cutting-edge and credible scientific knowledge has %introduced the exploitation of the demanding
made scientific literature part of the fact-checking sources.

Publishing fact-checked information has a positive effect on correcting false and inaccurate information, discouraging bad actors to spread misinformation. However,
%the corrections of the process may not stand the test of time
the fact-checked information does not necessarily prevail, due to the persistent promotion of less accurate claims from highly influential groups \cite{nyhan2021backfire}. The same is also true for the medical domain, where the decrease in COVID-19 misperceptions thanks to fact-checking, does not persist over time even after repeated exposure \cite{carey2022ephemeral}. As a result, fact-checked information is only part of the solution towards credible medical information, and it should be combined with the automatic fake-news detection and mitigation approaches discussed in the next sections to be effective.

The process of fact-checking is either done manually, by domain experts or workers in crowdsourcing platforms, or automatically.
%using knowoledge bases that are considered the authority.
The two approaches are discussed in detail below, while a detailed description of the available online fact-checking tools is provided in Section \ref{sec:tools}.

\subsection{Manual Fact-Checking}
\label{sec:manualFC}
In manual fact-checking, the assessment of a claim as true or false is done by people who read the articles that must be checked and decide whether the contained claims are true or false based on certain criteria, metrics, and research. Manual fact-checking is divided into expert-based and crowd-sourced-based fact-checking.

\subsubsection{Expert-Based Fact-Checking}
In expert-based fact-checking, the news is checked by experts of a domain, like the previously mentioned COVID-19 fact-checkers network. Usually, this team consists of journalists and domain experts. This method offers accurate fact-checking but is expensive and time-consuming. Moreover, it is difficult for a group of people to cover every day all the current affairs articles and keep up with rapidly changing domains like the COVID-19 pandemic.

The most famous fact-checking site that makes use of an expert-based fact-checking method and was highly active during the pandemic is Politifact \cite{Politifact}, which contains a column about health news and a column about coronavirus. In Politifact, a team of experts studies daily transcripts, news stories, press releases, and campaign brochures to find the most significant claims. Politifact uses the Truth-O-Meter ratings that classify the claims into the following categories: True (accurate), Mostly True (needs clarification), Half True (leaves out key details), Mostly True (ignores critical facts), False (not accurate), and Pants on fire (ridiculous claim).

Another platform is the Media Bias/Fact Check \cite{MediaBias} where the team uses a 0-10 scale to rate sites for biased wording, headlines, actuality, sourcing, story choices, and political affiliation. The team has sorted the various sources into the following bias categories: Left, Left-Center, Right-Center, Right, Least Biased, Conspiracy-Pseudoscience, Questionable Sources, Pro-Science, and Satire.

Based on the work of expert fact-checkers, a number of datasets have been constructed for facilitating research for claim detection and verification. A well-known generic fake news dataset is FEVER \cite{thorne2018fever}. It consists of almost 200K claims manually verified against Wikipedia pages and classified as SUPPORTED, REFUTED, or NOTENOUGHINFO.
Datasets that focus on assessing the veracity of scientific claims include the SciFACT \footnote{\url{https://github.com/allenai/scifact}} \cite{wadden2020fact} and the larger SciFACT-Open\footnote{\url{https://github.com/dwadden/scifact-open}} \cite{wadden2022scifact} datasets. These datasets include expert-written claims paired with evidence-containing abstracts annotated with veracity labels and rationales. Datasets that focus on the opposite task of non-claim detection also exist \cite{prabhakar2020claim}.  Finally, there exists a dataset\footnote{\url{https://borealisdata.ca/dataset.xhtml?persistentId=doi:10.5683/SP2/VPYSIS}} containing COVID-related claims that have been reviewed by fact-checking organizations around the world and retrieved by the Google Fact Check Tools API\footnote{\url{https://toolbox.google.com/factcheck/explorer}}.

\subsubsection{Crowd-Source-Based Fact Checking}
In crowd-sourced fact-checking, the detection of fake news is done by a large population that rates the credibility of articles. This approach has been proposed by various organizations such as WikiTribune \cite{WikiTribune} and is much more economical than expert-based fact-checking. However, it carries the risk that the rating population might introduce their own biases in the process.

The process consists of several steps. The first step is monitoring the news on TV, social media, newspapers, and websites and selecting the articles to be checked. When this selection is made by experts, the articles are filtered and balanced in order to be unbiased (e.g., for covering the whole political spectrum). This is a major differentiation from the case where the selection of articles is made up by the crowd since in the latter case it is difficult to certify that the selection of new claims is spread fairly across the news spectrum. The next step is researching the subject of the claims using multiple sources and
%the conclusion of the fact-checking procedure.
assigning a rating to the news article. This step is difficult to be completed objectively by people who are not experts. There are concerns about the correctness of the rating, the availability of evidence, and the rater's motivation. Another drawback of this approach is that it is difficult for volunteers to fact-check a claim that needs to be assessed rapidly, as in the case of the pandemic.

An example of a site that works with crowd-sourced fact-checking is Fiskkit \cite{Fiskkit}, where users can select articles, upload them to the site and rate them sentence-by-sentence. Users can also apply tags that evaluate the article’s accuracy and see the ratings of other users.

Regarding available datasets, the Multi-Genre Natural Language Inference Corpus MNLI\footnote{\url{https://cims.nyu.edu/sbowman/multinli/}} is a crowd-sourced collection of 433K sentence pairs annotated for textual entailment. In other words, the data consists of pairs (p, h), where p is the premise and h is the hypothesis, and labels in \{entailment, contradiction, neutral\}, which report whether the hypothesis entails, contradicts or is neutral towards the premise.

\subsection{Automatic Fact-Checking}
\label{sec:automaticFC}
Manual fact-checking has satisfactory results, especially in the case of expert-based fact-checking. However, since the task of fact-checking is time-consuming, this method is not efficient for the rapidly changing domain of news, and for keeping pace with the volume of new content on the web.
%Therefore, the scientific community, and especially the artificial intelligence (AI) community, has been exploring approaches to automate fact-checking, by exploiting techniques and advancements from the domains of natural language processing (NLP), sentiment analysis (SA), machine learning (ML), and deep learning (DL), and applying them to the various unstructured (posts, comments, etc.) and structured data (graphs and knowledge bases).
Therefore, the scientific community has been exploring approaches to automate fact-checking by exploiting techniques and advancements from the domains of NLP and DL. The process of automatic fact-checking consists of two steps: a) detecting the claims in the text and b) assessing the validity of the claims by retrieving evidence.

\subsubsection{Claim Detection}
Regarding the detection of the claims contained in social media posts, comments, news, and web pages, it is important that the most check-worthy claims are selected. Such claims are those for which people show interest and are trending. The claims are usually collected from social media, where metadata like the number of likes and reposts are used as features to identify top claims \cite{di2022health}.  Other sources of claims can be found in Wikipedia (e.g., COVID-19 pandemic misconceptions) or in news websites and organizations.

In \cite{wuhrl2021claim} the authors annotate a corpus of 1200 tweets for implicit and explicit biomedical claims. Using this corpus, which is related to COVID-19, measles, cystic fibrosis, and depression, they developed deep-learning models that automatically detect tweets containing claims. Their analysis showed that biomedical tweets are densely populated with claims.  Despite the fact that the detection of claims was challenging, they report that deployed models provided acceptable performance.

While most works focus on single claim sentence analysis, the work presented in \cite{reddy2022newsclaims} introduces the NewsClaims dataset, a benchmark for attribute-aware claim detection considering topics related to COVID-19. Specifically, given a news article, the task is to identify the claim sentence to a set of predefined topics that contain factually verifiable topics, the claimer, the claim object, the stance of the claimer, and the exact claim boundaries. For claim sentence detection they use Claimbuster\footnote{\url{https://idir.uta.edu/claimbuster/api/}} \cite{hassan2017claimbuster} along with pre-trained Natural Language Inference (NLI) models as zero-shot classifiers based on BART, where the claim sentence is the NLP premise and the hypothesis is constructed from each topic. The claim object task is modeled as a zero-shot or few-shot setting by converting it into a prompting task for pre-trained language models like GPT-3\footnote{\url{https://lablab.ai/tech/gpt3}}. Stance detection is done again through NLI, where the affirm and refute labels construct the hypothesis,  taking as stance the corresponding higher entailment score. Claim boundary detection is done using fine-tuned Bidirectional Encoder Representations from Transformers (BERT)\cite{devlin2018bert} models, the Project Debater APIs \cite{bar2021project} and the PolNeAR\footnote{\url{https://github.com/networkdynamics/PolNeAR}} popular news attribution corpus of annotated triples comprising the source, cue, and content for statements made in news. Finally, regarding claimer detection, they again leverage PolNeAR for building a claimer extraction baseline by fine-tuning a BERT model, along with a second baseline built upon Semantic Role Labeling (SRL), that outputs the predicate-argument structure of a sentence such as who did what to whom. The results showcase that the above tasks, except the task of stance detection, are difficult for current models, especially the task of claim sentence detection.

\subsubsection{Claim Validation}
For the assessment of the validity of a claim, an essential part is the retrieval of the evidence process. Evidence retrieval is the task of retrieving documents that support the prediction of a claim. During this process, information and proofs must be found around the claim, such as text, tables, knowledge bases, images, and other metadata for evidence of the truth.
%A fundamental issue is that not all the available information is trustworthy. Most fact-checking approaches make use of trustworthy sources (e.g., encyclopedias).
A fundamental issue is finding trustworthy sources. For example, many fact-checking approaches make use of encyclopedias.
Table \ref{tab1} reports some trustworthy sites for information and proof of medical claims.

After the retrieval of evidence, a fact-verification method has to conclude the validity of a claim. Usually, the verification of claims leverages NLI techniques \cite{storks2019recent}. As already mentioned, the NLI task aims to classify the relationship between a pair of a premise (evidence) and a hypothesis (claim) as either entailment, contradiction, or neutral. However, in fact-verification systems, the usually multiple evidence pieces, are found by the systems themselves. In addition, given a collection of false/true claims, the verification of a new information piece can also be modeled as a NLI task, where the goal is to detect entailment with one of the false/true collected claims \cite{martin2022facter}.  Another approach is described in \cite{wright2022generating}, where the authors present the ClaimGen-BART and Knowledge Base Informed Negations (KBIN) methods for generating claims and claim negations  supported by the literature, using the BART pre-trained model \cite{lewis2019bart}.
In \cite{wadden2020fact} the authors present  Kernel Graph Attention Network (KGAT), which conducts more fine-grained fact verification with kernel-based attention.

\begin{footnotesize}
\begin{table}
\caption{Trustworthy websites for medical news}
\label{table}
\begin{tabular}{|l|l|}
\hline
\textbf{Name} &  \textbf{URL}  \\\hline\hline
World Health Organization &  \url{https://www.who.int/}  \\\hline
Medscape &  \url{https://www.medscape.com/}  \\\hline
News-Medical Life Sciences &  \url{https://www.news-medical.net/}  \\\hline
Medgadget &  \url{https://www.medgadget.com/}  \\\hline
Medical News Today &  \url{https://www.medicalnewstoday.com/}  \\\hline
Medical Xpress &  \url{https://medicalxpress.com/}  \\\hline
The BMJ &  \url{https://www.bmj.com/}  \\\hline
Stanford Medicine &  \url{https://med.stanford.edu/}  \\\hline
MedlinePlus &  \url{https://medlineplus.gov/}  \\\hline
Healio &  \url{https://www.healio.com/}  \\\hline
Centers For Disease Control and Prevention &  \url{https://www.cdc.gov/}  \\\hline
Modern Healthcare &  \url{https://www.modernhealthcare.com/}  \\\hline
\end{tabular}
\label{tab1}
\end{table}
\end{footnotesize}

Regarding the medical domain, a COVID-19-specific dataset that has been constructed using automatic methods is COVID-Fact \cite{saakyan2021covid} which contains 4,086 claims concerning the pandemic, evidence for the claims, and contradictory claims refuted by evidence. The approach described in \cite{liu2020adapting} adapts the open-domain fact extraction and verification KGAT approach\cite{liu2019fine} with in-domain language models, based on the SciFACT and COVID-Fact datasets. Specifically, the in-domain language model transfers COVID domain knowledge into pre-trained language models with continuous training. The COVID medical token semantics are learned using mask language model-based training.
In a similar manner, the authors of \cite{kotonya2020explainable} introduce the PubHealth dataset for public health fact-checking, which also includes explanations, and explore veracity prediction and explanation generation tasks using various pre-trained models. Their results show that training
%veracity prediction and explanation generation
models on in-domain data improves the accuracy of veracity prediction and the quality of generated explanations compared to training generic language models without explanation.

In \cite{sarrouti2021evidence} the authors introduce the HealthVer dataset for evidence-based fact-checking of health-related claims. The dataset was created using a three-step approach. The first step is to retrieve real-world claims from snippets returned by a search engine for questions about COVID-19. The next step is to retrieve and rank relevant scientific papers as evidence from the COVID-19 Open Research Dataset (CORD-19) \cite{wang2020cord}  using a T5\footnote{\url{https://github.com/google-research/text-to-text-transfer-transformer}} relevance-based model. The last step is to manually annotate the relations between each evidence statement and the associated claims. The conducted experiments showed that training deep learning models on real-world medical claims greatly improves performance compared to models trained on synthetic and open-domain claims.

In \cite{wadhwa2022redhot} the authors evaluate baseline models for categorizing reddit posts as containing claims, personal experiences, and/or questions. In addition, they evaluate various BERT models for extracting descriptions of populations, interventions, and outcomes, as well as for tagging claims, questions, and experiences. Finally, using snippets they retrieve trustworthy (published) evidence relevant to a given claim. To this end, they introduce a heuristic supervision strategy that  outperformed pre-trained retrieval models.

\subsubsection{Knowledge Bases and Claim Validation}
Another direction for assessing the validity of a claim and retrieving the corresponding evidence is by exploiting knowledge bases. A knowledge base (KB) is a collection of information and resources where human knowledge can be stored. A common way of representing it is by connecting two entities with a given relationship. These relationships can form a graph, the knowledge graph (KG), where the entities are represented as nodes and relationships are represented as edges. In this case, a fact is defined as a triple that has the form of ("subject" s, "predicate" p, "object" o), and can be classified to different categories (e.g., numerical, object properties, etc.). Usually, the data modeling languages that are used for creating the graphs are the Resource Description Framework (RDF)\footnote{\url{https://www.w3.org/RDF/}} or the more expressive Web Ontology Language (OWL)\footnote{\url{https://www.w3.org/OWL/}} that also provides automatic reasoners. An example is KG-Miner \cite{shi2016discriminative} which can predict the truthfulness of a statement using discriminative predicate path mining.

Given a textual claim and a KG,  the claim is converted to a triple by using NLP methods \cite{huynh2018towards,angeli2015leveraging}, and the validity of the fact is checked against the information contained in the graph. However, the KG is considered incomplete (Open World Assumption), meaning that
%if it was complete any fact could be easily validated by existentially checking whether the statement is contained in the graph or not.
it does not contain all known true facts.
As a result, a missing fact does not imply an invalid claim. The proposed methods in the literature try to overcome this issue.

There are three main approaches for checking the validity of a claim against an incomplete KB \cite{kim2021fact}: a) using external web resources as a way to find new fact triples that are missing from the KB and complement existing knowledge \cite{distiawan2019neural}, b) embedding-based \cite{nguyen2017novel, meilicke2018fine}  and path-based approaches \cite{shi2016discriminative} that use graph embeddings and  properties of the paths as features respectively for verifying facts, and c) rule-based approaches that use rule-mining techniques to validate a fact \cite{shiralkar2017finding}. The first approach, which complements the KB with external knowledge, has been proven to be inaccurate due to the difficulty of the information extraction task (83.51\% of fact-triples extracted from Wikipedia using a BERT-model relation extraction approach were false \cite{kim2021fact}). Embedding-based approaches, since they use a statistical approach, have the benefit that they can verify entity pair links that are not linked in the KG, covering more verifiable triples which are not verifiable by rule-based ones \cite{kim2021fact}. On the other hand, since rule-based approaches use logical rules, their results are more interpretable, and they can easily verify some facts that the embedding approaches have trouble with \cite{meilicke2018fine}. Finally, it has been shown that both the embedding-based and rule-based approaches can be used complementarily, offering better performance than by using them separately \cite{kim2021fact}.

In the medical domain, there are various knowledge graphs that could be used for fact-checking claims, such as COVID-19, CovidGraph\footnote{\url{https://healthecco.org/covidgraph/}}, literature-review related ones \cite{chatterjee2021knowledge}, oncological-specific KGs \cite{silva2022ontologies}, personalized medicine recommendation KBs \cite{chandak2022building, silva2022matching}, disease ontologies like ICD-9\footnote{\url{https://bioportal.bioontology.org/ontologies/HOM-ICD9}} and drug safety and interactions KBs like DrugBank\footnote{\url{https://go.drugbank.com/}}.

The SciClaim KG \cite{magnusson2021extracting}
is a graph of scientific claims drawn from Social and Behavior Science (SBS), PubMed, and CORD19 papers. It incorporates coarse-grained entity spans (list of tokens) as nodes and relations as edges between
them, and fine-grained attributes that modify entities and their relations. In total it contains
12,738 labels that capture causal, comparative, predictive, statistical, and proportional associations over experimental variables, along with their qualifications, subtypes, and evidence. The schema is inferred using a transformer-based joint entity and relation extraction approach.

The UMLS\footnote{\url{https://www.nlm.nih.gov/research/umls/index.html}} meta-thesaurus is a large biomedical knowledge base that unifies hundreds of different ontologies in biomedicine. UMLS is used as the source knowledge base for normalization and candidate selection for KBIN. Additionally, it is the knowledge base used to train the clinical concept embeddings cui2vec\footnote{\url{https://github.com/beamandrew/cui2vec}} \cite{beam2019clinical}, which are used for candidate concept selection in KBIN.

The public medical knowledge graph KnowLife \footnote{\url{http://knowlife.mpi-inf.mpg.de/}} \cite{ernst2015knowlife} contains 25,334 entity names and 591,171 triples, and it is used among others by \cite{cui2020deterrent} to extract six positive relations (i.e., Causes, Heals, CreateRiskFor, ReducesRiskFor, Alleviates, and Aggravates) and four negative relations (i.e., DoesNotCause, DoesNotHeal, DoesNotCreateRisk, and DoesNotReduceRiskFor). The approach uses an attention mechanism to calculate the importance of entities for each article and the knowledge-guided article embeddings are used for misinformation detection.

A remark about the above methods is that although we assume that KBs and KGs contain only accurate information, this is usually not true. For example, Wikipedia, which is used as a source in the construction of various KGs, is known to contain various false facts in different time snapshots. For example, version 3.6 of the DBpedia KG, which is constructed using Wikipedia, has been found to be 80\% correct \cite{gerber2015defacto}, while for most KGs this accuracy is unknown.

\section{Content-based Methods}
\label{sec:content}
Content-based fake news detection methods analyze the content of the various sources and their interactions with it \cite{shu2019defend}. Specifically, they extract various features from news items, social media posts, user comments, or the content of external sources through hyperlinks and quotes \cite{sym13061091}.
Before feature extraction, it is necessary to pre-process the data by using techniques such as tokenization, lowercase transformation, removal of stop words, sentence segmentation, etc. Afterwards, the extracted features can be used as the input to ML classifiers or can be concatenated or aggregated in neural network architectures like feedforward neural networks (FNNs), recurrent neural networks (RNNs), convolutional neural networks (CNNs), Long Short-Term Memory networks (LSTMs), Generative Adversarial Networks (GANs), Sequence-to-Sequence networks (Seq2Seq) like transformers, Graph Neural Networks (GNNs), etc. (see \cite{10.1145/3533431}, \cite{data7050065}, \cite{9599970}, \cite{10.1145/3548458}, \cite{9350542}).

Below we describe the different types of content-based features commonly used in the literature. We consider the following categories: text representations, linguistic features, emotional features, entity-based features, stylistic features, topic extraction, user-profile features, image-based features, and external features. Table \ref{table:contentBased} offers an overview of all the available features.

%Feature extraction is important to provide useful information to ML methods. Features can be separated into general features and latent features.

\subsection{Text representations}
Text representations address the fundamental problem of converting unstructured documents to mathematically computable forms, usually in the form of vectors. Commonly used approaches include the bag-of-words (BoW) approach, which weights the frequency of a word in a document \cite{di2022health,bangyal2021detection}, and the term-frequency-inverse-document-frequency (TF-IDF) approach, which keeps the appearances of words in a document, by the fraction of documents in the corpus that contain the word\cite{di2022health,bangyal2021detection,dhoju2019differences,pelrine2021surprising}. Approaches using n-grams,  which are based on contiguous sequences of n 'items' (words, letters, syllables, phonemes, etc.), are commonly used even in the context of medical fake news detection \cite{pelrine2021surprising}.

More modern approaches use a latent space to create text embeddings over the content of the articles. Embeddings provide a way to translate a high-dimensional vector into a relatively low-dimensional one, placing semantically similar inputs close together in the embedding space. Usually, embeddings are learned over big corpora and datasets, so that they can be reused across models, domains, and tasks. It has been shown that the effectiveness of word embeddings in encapsulating the semantic, syntactic, and context features affects the performance of the classification models \cite{truicua2023s}.

Text embeddings can be divided into non-contextual and contextual ones. Non-contextual ones, like word2vec \cite{mikolov2013efficient}, GloVe \cite{pennington2014glove}, and FastText\cite{joulin2016fasttext}, learn word associations from a large corpus of text but do not capture information about the context in which they appear. They provide a single global representation for every word, even if words can be ambiguous and have various meanings. For example in the context of medical fake news,  the authors in \cite{data7050065} use the GloVe pre-trained model to classify COVID-19 news by using various neural network architectures.

Contextual embeddings assign to each word a representation based on its context. In this way, they can capture uses of words across varied contexts and can encode knowledge that can be cross-lingual. In \cite{truicua2023s}, the authors evaluate non-contextual and contextual text embeddings approaches over various fake-news datasets, and showcase the importance of contextual embeddings. A large number of approaches that leverage latent features use pre-trained language models based on the popular BERT model \cite{devlin2018bert} and its derivatives (BART \cite{lewis2019bart}, RoBERTa \cite{liu2019roberta}, distilBERT \cite{sanh2019distilbert}, Albert \cite{lan2019albert}, XLNet \cite{yang2019xlnet}, etc.). Such models can be fine-tuned over a collection of fake and valid news for classification reasons, as is the case of FakeBERT \cite{kaliyar2021fakebert} and offer exceptional performance for the task at hand.
Furthermore, they can be trained over datasets of a specific domain, in order to identify the terminology, constructs, and peculiarities of this domain.
An example of a pre-trained representation model in the medical domain is BioBert \cite{lee2020biobert}, a BERT model for Biomedical Text Mining. This is the first domain-specific language representation model pre-trained on large-scale biomedical corpora (PubMed abstracts and PMC full-text articles). The reported results show that pre-training BERT on biomedical articles is crucial for several tasks in the medical domain. In the same manner, Med-BERT \cite{rasmy2021med} which has been trained in electronic health records, has shown excellent performance in disease prediction tasks on two clinical databases. Another relevant BERT-based model is the scientific-BERT (SCIBERT) \cite{kumari2021debunking}, which was trained on the FakeHealth dataset that contains health-related annotated news from the news fact-checking site Snopes\footnote{https://www.snopes.com/}. Finally, BioGPT \cite{luo2022biogpt} is a domain-specific generative Transformer language model pre-trained on large-scale biomedical literature.

\begin{footnotesize}
\begin{table}[h!]
\caption{Pre-trained models: General, fake-news, medical domain, and social media-related ones over various datasets}
\centering
\begin{tabular}{l  l  r }
 \multicolumn{3}{c}{\textbf{General}}\\\hline
 BERT & English Wikipedia, BooksCorpus  &  \cite{devlin2018bert}\\
BART & English Wikipedia, BooksCorpus & \cite{lewis2019bart} \\
RoBERTa & English Wikipedia, BooksCorpus, & \cite{liu2019roberta} \\
  &  CC-News, OpenWebText,   & \\
  &  and CommonCrawl Stories  & \\
  distilBERT & English Wikipedia, BooksCorpus & \cite{sanh2019distilbert} \\
  Albert & English Wikipedia, BooksCorpus & \cite{lan2019albert} \\
  XLNet & English Wikipedia, BooksCorpus & \cite{yang2019xlnet} \\
  \\
\multicolumn{3}{c}{\textbf{Fake News}}\\\hline
 FakeBERT & 40K fake and valid news& \cite{kaliyar2021fakebert} \\
 \\
 \multicolumn{3}{c}{\textbf{Medical Domain}}\\\hline
cui2vec & 20M clinical notes and & \cite{beam2019clinical}\\
 & 1.7M biomedical articles & \\
BioBert & PubMed abstracts and PMC articles & \cite{lee2020biobert} \\
Med-BERT & Electronic health records &  \cite{rasmy2021med} \\
SCIBERT & FakeHealth dataset (news) &  \cite{kumari2021debunking} \\
COVID-Twitter-BERT & 22.5M COVID-19  tweets &  \cite{DBLP:journals/corr/abs-2005-07503}\\
BioGPT & 15M PubMed abstracts & \cite{luo2022biogpt} \\
 \\
\multicolumn{3}{c}{\textbf{Social Media}}\\\hline
BERTweet & 850M English Tweets  &  \cite{DBLP:journals/corr/abs-2005-10200} \\
\end{tabular}
\label{table:pretrainedModels}
\end{table}
\end{footnotesize}

Other strategies for fake news detection \cite{De2022MultiContextBN, DBLP:journals/corr/abs-2101-02359, DBLP:journals/corr/abs-2101-00180, 9418446, https://doi.org/10.48550/arxiv.2208.01355, 9803414} use a combination of different types of pre-trained models over different domains and contexts. For example, in \cite{De2022MultiContextBN} authors propose a multi-context neural architecture with three pre-trained transformer-based models: BERT, BERTweet \cite{DBLP:journals/corr/abs-2005-10200} and COVID-Twitter-BERT \cite{DBLP:journals/corr/abs-2005-07503}. They show that their combination outperforms the use of a single model. The reason is that the hybrid model benefits from the different aspects that each model understands. Specifically,  BERT understands the English language constructs, BERTweet the structures of tweets, and COVID-Twitter-BERT the COVID-19 scientific terms.

Finally, several knowledge embedding models have been widely proposed in the bibliography. Such models include  TransE \cite{bordes2013translating}, TransD \cite{ji2015knowledge}, TransMS \cite{yang2019transms} and TuckER \cite{balavzevic2019tucker}.  The performance of the above models has been explored in the context of Twitter COVID-19 misinformation detection, as a graph link prediction problem in \cite{weinzierl2021automatic} and \cite{weinzierl2022identifying}. In \cite{cui2020deterrent} medical knowledge graph embeddings are leveraged to guide article embedding using a graph attention network to  differentiate misinformation from fact.

An important conclusion from the plethora of the aforementioned works is that exploiting pre-trained models benefits studies on other smaller or domain-specific datasets, by offering exceptional performance in various tasks, while at the same time reducing data collection costs. Table \ref{table:pretrainedModels} summarizes the mentioned pre-trained models.

\begin{footnotesize}
\begin{table*}[h!]
\centering
\caption{Features used in content-based approaches}
\begin{tabular}{| l | l | l |}
 \hline
 Feature Type & Description & Works \\
 \hline\hline
 \multirow{5}{*}{Textual  Representations}  & BoW & \cite{di2022health,bangyal2021detection}\\
      & TF-IDF & \cite{di2022health,bangyal2021detection,dhoju2019differences,pelrine2021surprising, DBLP:journals/corr/abs-2101-00180}\\
      & n-grams & \cite{pelrine2021surprising} \\
      & Non-Contextual embeddings & \cite{data7050065, DBLP:journals/corr/abs-2101-02359, di2022health, pelrine2021surprising,dharawat2022drink, truicua2023s, endo2022illusion}\\
      & Contextual embeddings& \cite{liu2020adapting,baris2021ecol,kar2021no,baris2021ecol,pelrine2021surprising,dharawat2022drink, truicua2023s, De2022MultiContextBN, DBLP:journals/corr/abs-2101-02359, DBLP:journals/corr/abs-2101-00180, 9418446, https://doi.org/10.48550/arxiv.2208.01355, 9803414, shu2019defend}\\
      & Knowledge embeddings & \cite{cui2020deterrent,weinzierl2021automatic,weinzierl2022identifying}\\
  \hline
 \multirow{5}{*}{Linguistic}  & PoS & \cite{spalenza2021lcad, di2022health, pelrine2021surprising,dai2020ginger}\\
    & Uncertainty constructs &\cite{dhoju2019differences,DBLP:journals/corr/abs-1809-00557} \\
    & Certainty constructs  &  \cite{dhoju2019differences}\\
    & Passive voice & \cite{park2013content} \\
    & Headline length & \cite{dhoju2019differences}\\
 \hline
 \multirow{4}{*}{Emotional}  & Sentiment analysis & \cite{mukherjee2014people,salvi2021going,di2022health, iwendi2022covid,khan2022detecting,solovev2022moral}\\
    & Subjectivity & \cite{endo2022illusion}\\
    & Polarity  & \cite{iwendi2022covid, mukherjee2014people}\\
 \hline
 \multirow{4}{*}{Entities}  & NER & \cite{spalenza2021lcad, di2022health}\\
    & Medical terms & \cite{di2022health}\\
    & Commercial terms & \cite{di2022health}\\
    & Hyperlinks & \cite{dhoju2019differences} \\
    & Hashtags & \cite{zhou2022fake} \\
    & Direct quotes & \cite{dhoju2019differences} \\
    & Visual and textual entities inconsistencies& \cite{qi2021improving}\\
 \hline
  \multirow{3}{*}{Stylistic}  & Information, Complexity, Readability, Specificity, & \\
   & \& Informatility, Non-immediacy, Diversity & \cite{endo2022illusion,santos2020measuring,bangyal2021detection,DBLP:journals/corr/abs-1809-00557}\\
    & Misspellings & \cite{bangyal2021detection}\\
    \hline
 \multirow{2}{*}{Topic Extraction} & LDA & \cite{sabeeh2021fake} \\
  & Markov-chains & \cite{ceron2021fake} \\
 \hline
 \multirow{1}{*}{Image}  & Visual features  &\cite{qi2021improving, shang2022duo, uppada2022image} \\
  \hline
 \multirow{1}{*}{User Profiles}  & User description, timeline, friends profile & \cite{kar2021no, dai2020ginger}\\
  \hline
   \multirow{3}{*}{External} & Fact-verification score  & \cite{kar2021no} \\
    & Credibility score & \cite{baly2018predicting}\\
    & Stance Detection & \cite{umer2020fake, ghanem2018stance, peskine2021detecting} \\
  \hline
\end{tabular}
\label{table:contentBased}
\end{table*}
\end{footnotesize}

\subsection{Linguistic features}
Linguistic features are characteristics that are used to classify phonemes or words, along with their inter-relationships. Of special importance in the context of fake news detection are morphological, syntactical, and grammatical features like Part-of-speech (POS) and Noun-phrases (NPs) \cite{kansal2021fake, pelrine2021surprising, di2022health, dai2020ginger,bangyal2021detection}.
Various works showcase the importance of uncertainty constructs like possibility adverbs, conditional particles, and certainty constructs like demonstrative adjectives/pronouns, and declarative conjunctions. For example, the authors in \cite{dhoju2019differences} study the structural difference between fake and true news for the health domain. They claim that the size of the headline is a key element of a news article, and they analyze the presence of some common patterns like the use of demonstrative adjectives, numbers, modal words, questions, and superlative words. Similarly, the authors in \cite{park2013content} report that the content provided by trustworthy sources tends to be expressed in a passive voice. Finally, the use of morphological Part-Of-Speech (POS) n-grams was not found to improve fake news detection over a TF-IDF baseline \cite{kapusta2021using}.

\subsection{Emotional features}
Sentiment analysis and emotion detection have also been studied in the context of fake news detection \cite{mukherjee2014people,salvi2021going,di2022health, iwendi2022covid,endo2022illusion,khan2022detecting,solovev2022moral,iwendi2022covid, bangyal2021detection}. Sentiment analysis characterizes textual content as positive, negative, or neutral, while emotional detection identifies distinct human emotion types like depression, happiness, anger, etc.
Fake news has been noted for producing feelings of fear and anxiety to news consumers \cite{salvi2021going}. In the same manner, a high level of negativity and depression has been correlated with subjectivity \cite{mukherjee2014people}. The polarity of the text has also been used for classifying fake and valid news since polarity indicates bias and questionable credibility, and can be extracted using emotion analysis techniques \cite{iwendi2022covid,mukherjee2014people}.

There is a plethora of tools used for both sentiment detection and emotional analysis. Such resources include the NRC lexicon\footnote{\url{https://saifmohammad.com/WebPages/NRC-Emotion-Lexicon.htm}}, the MPQA Subjectivity lexicon\footnote{\url{https://mpqa.cs.pitt.edu/}}, the Affective Norms for English words (ANEW) lexicon \footnote{\url{https://github.com/nisarg64/Sentiment-Analysis-ANEW}}, the SentiWordNet sentiment lexicon\footnote{\url{https://github.com/aesuli/SentiWordNet}}, the WordNet-Affect affective concept and words lexicon\footnote{\url{https://wndomains.fbk.eu/wnaffect.html}},
the text2emotion python package\footnote{\url{https://pypi.org/project/text2emotion/}}, the general NLP task TextBlob library \footnote{\url{https://textblob.readthedocs.io/en/dev/}}, and others.

\subsection{Entity-based features}
News usually contains named entities which can be important for understanding the news semantics and for detecting fake news. Such entities include categories like Person, PersonType, Location, Organization, Event, Product, Dates, and others. The process of identifying the named entities is called Named Entities Recognition (NER)\footnote{A list of NER tools is provided in \url{https://www.clarin.eu/resource-families/tools-named-entity-recognition}}. Biomedical NER has also been studied for recognizing diseases, drugs, surgery reports, anatomical parts, and examination documents \cite{ramachandran2021named, wen2021medical, zhou2021end}. In addition, the previously presented language models  can be fine-tuned for this specific task \cite{sharma2019bioflair, lee2020biobert}.

NER approaches are usually deployed in fake news classification, as a feature that enriches the textual representation, sentiment, and linguistic features \cite{spalenza2021lcad, bangyal2021detection}. For example, \cite{spalenza2021lcad} uses a combination of POS tagging and NER sequences to identify valid and fake news. The work described in \cite{di2022health} uses Linguistic-Medical entities and specifically the fraction of medical terms in a document. Those are found using the MedMentions\footnote{\url{https://github.com/chanzuckerberg/MedMentions}} NER model, which is specially trained on medical information, and consists of 4,392 biomedical papers annotated with mentions of UMLS entities. They also exploit commercial terms since they are used for profit. As a result the higher the number of commercial terms the less credible the information \cite{freeman2004examination}.
%For dealing with the out-of-date problem with the medical development or emergence of a new virus, the authors in \cite{dhoju2019differences} propose continuous in-domain training with the latest medical corpus. In addition,
The authors in \cite{dhoju2019differences} explore to what extent credible medical-related sources make use of quotations and hyperlinks in a news article, by using the Stanford QuoteAnnotator\footnote{\url{https://github.com/muzny/quoteannotator}}. What they observed is that unreliable outlets use a smaller number of quotations compared to reliable ones. Regarding the hyperlinks, they concluded that a credible news article contains on average 1.6 more hyperlinks than non-credible articles. They also compare various classifiers using textual representations, headline length, number of direct quotes, and number of hyperlinks and showcase the importance of direct quotes and hyperlinks for detecting fake news. Hashtags have also been found effective in identifying fake news, especially early in the propagation life-cycle \cite{zhou2022fake}. Finally, a multimodal approach  that monitors inconsistencies between recognized entities in textual and image content, offering a potential indicator for multimodal fake news, is described in \cite{qi2021improving}.

\subsection{Stylistic features} Various stylistic capturing features of the content are considered to be disinformation-related  by \cite{zhou2020survey} and are  studied in various works \cite{endo2022illusion,santos2020measuring,bangyal2021detection}. Such attributes are related to the quantity of information present in the text, its complexity (e.g., the ratio of characters and words in a message), its non-immediacy (deceivers tend to disassociate themselves from their deceptive messages for reducing accountability and responsibility for their statements \cite{zhou2004comparison}), its diversification in terms of words,
its uncertainty (e.g., lack of information, hedging, lexical ambiguity, and negation), its informality (e.g., presence of emoticons, slang or colloquial words), its specificity (e.g., deciding when to convey general
statements, elaborate on details, and gauging how much details to convey), and its readability (accessibility of a text showing how wide an audience it will reach) \cite{santos2020measuring}. Misspellings in a COVID-19 setting have been studied in
\cite{bangyal2021detection}.

\subsection{Topic extraction}
Topic extraction methods have also been used in fake news detection approaches. Specifically, \cite{sabeeh2021fake} uses the Latent Dirichlet allocation (LDA) \cite{blei2003latent} topic detection model to examine the influence of the article’s headline and the body, both individually and collaboratively. These features are then used to enhance a BERT-based classifier. A Markov-inspired topic extraction approach has been used for identifying trends in fake news agendas through fact-checking content over a time window \cite{ceron2021fake}.

\subsection{User-profile features}
The content of a user profile along with the user timeline are also important for fake news detection.
For example, in \cite{kar2021no} the authors use the textual information of the user description along with other properties of the user account. In addition, the user timeline and the profile of the user friends have been studied \cite{dai2020ginger}. We discuss in more detail the user-related features in Section \ref{sec:sources}.

\subsection{Image-based features}
Posts on social media are frequently accompanied by images or even sometimes consist of just plain images (e.g., memes). Image-containing posts generally target the emotions of the people, spread faster, and have a high level of retweets and shares. If images contain text (e.g., in the caption or inline), the previously discussed approaches can be applied in the extracted text. Else more complicated deep learning pipelines \cite{uppada2022image} can be deployed, that flag fake posts with visual data embedded with text. Such approaches can even support the identification of tampered and out-of-context images. The already discussed work described in \cite{qi2021improving} tries to identify inconsistencies between recognized entities in textual and image content. Finally, \cite{shang2022duo} proposes a multimodal fused encoder-decoder architecture to detect COVID-19 misinformation by investigating the cross-modal association between the visual and textual content that is embedded in the multimodal news content. The challenging part of this method is to accurately assess the consistency between the visual and textual content. The deployed architecture uses: a) a multimodal feature encoder for the content (text, image) and user comments of input news articles, b) a text-guided visual feature generator that is given features from the previous encoder and generates visual features based on the understanding of the news text, c) an image-guided textual feature decoder to learn the corresponding guided textual feature information from the news images, and d) a comment-driven explanation generator that leverages both the original and generated features to provide content and comment explanations on the results.

\subsection{External}
Finally, another category of works tries to incorporate external features using textual information. For example, in \cite{kar2021no} the text of tweets is used as a query in a popular web search engine to compute a fact-verification score. Specifically, the authors consider as reliable the results of popular news websites only and use the Levenshtein distance between the tweet text and the titles of the results to compute the fact-verification score. Another approach \cite{baly2018predicting},  uses Wikipedia and Twitter among others, to assess the credibility of web pages. In the same manner, the provided references to external resources (e.g., URLs) can be used in combination with other features (e.g., sentiment, linguistic), to identify the credibility of the referenced information, by exploiting  available fact-checking services like the Media Bias Fact Checking service\footnote{https://mediabiasfactcheck.com/about/} and stance techniques \cite{umer2020fake, ghanem2018stance}.
%Pre-trained language models are also fine-tuned for stance detection using the NLI task over tweets\cite{peskine2021detecting}.
%, where the models are trained to classify tweets for entailment (agreement or support), contradiction (disagreement), or neutrality (undetermined) of claims \cite{peskine2021detecting}.
%We provide more details about the fake news sources in \ref{sec:sources} and the corresponding tools in \ref{sec:tools}.

\if false
\begin{table}[h!]
\caption{ML and DL algorithms/techniques}
\centering
\begin{tabular}{| c | c |}
 \hline
 Methods & Publication \\
 \hline
 \multicolumn{2}{|c|}{Classification}\\
 \hline
 Gradient Boosting & \\
 Logistic Regression & \\
 LDA & \\
 Naive Bayes & \\
 Random Forests &\\
 k-Nearest Neighbors & \\
 Decision Trees & \\
 SVM & \\
 SGD & \\
 MLP & \\
 CNN & \\
 RNN & \\
 Bi-directional RNN & \\
 LSTM & \\
 Bi-directional LSTM & \\
 GRU & \\
 Hierarchical Attention Networks & \\
 \hline
 \multicolumn{2}{|c|}{Claim Extraction}\\
 \hline
\end{tabular}
\label{table:pretrainedModels}
\end{table}
\fi

\section{Propagation-based methods}
\label{sec:prop}
Propagation-based methods exploit the several networks involved in and affecting the propagation of news in a social network. These methods require observing the behavior of users on the network, the characteristics of the users along with the contacts they interact with, and studying the patterns of the spread of fake news.

An important network is the \textit{cascade network} that captures the actual propagation of a news item in a social network. The  cascade network is usually a tree whose nodes represent the users who are involved in spreading each news item. The root represents  the publisher of the item, i.e, the user who first posted the item in the social network. There is a link from a node $u$ to a node $v$, if the user represented by $u$ has spread the item posted by the user represented by $v$, for example, by retweeting, reposting, or liking the item. Spreading news in such networks indicates endorsement of the original item. Note that for a specific news item, there may be more
than one publishers and thus multiple simultaneous cascades.

Another network directly capturing the propagation of an item is the \textit{discussion} network. This is usually a tree whose nodes represent items related to the news item.
The root represents the original news item and  there is a link for a node $u$ to node $v$, if the item represented by $u$ is a reply, or a comment to the item represented by node $v$. A common variation of the discussion network is one whose nodes represent the users that posted the items instead of the item itself. Spreading items in such networks indicates engagement with the original item but not necessarily endorsement of it, since a reply, or a comment may express a negative reaction.

Several other networks have also been studied, for example, networks that capture a friendship, or a follow relationships between the users spreading the news. We present some of them later in this section.

Propagation networks find many applications in the context of medical fake news. First, they have been exploited towards enhancing our understanding of the mechanisms driving the spread of fake news in a social network. Then, propagation networks are used to improve fake news detection.
Furthermore, they help in building richer user and item profiles. For example, a friendship network can be used to build an enhanced profile for a user $u$ that also incorporates information about the friends of $u$, or, a discussion network can be used to augment the content of an item $i$ with the content of replies to $i$.
This can be achieved either through explicit aggregation, or implicitly by using the propagation network to build for example a GNN architecture.

In the following, we first present propagation-based approaches that involve a feature extraction process. Then, we present approaches that use latent features of the networks.

\subsection{Propagation network features}
Previous research looks into several features of the networks involved in the propagation of the news items.
Such features have been used to make important observations regarding the spread of fake news
\cite{vosoughi2018spread,Safarnejad20}.
They have also been used as features for fake news detection \cite{di2022health,wu2015false}.
In Table \ref{tab:prop-features}, we report several features of propagation networks studied in the context of fake news \cite{Shu20,di2022health,Safarnejad20,vosoughi2018spread,Castillo11,Safarnejad20,ceylan2023sharing}.
Note that features can be computed not only at the whole graph, but also at specific subgraphs.

The most common \textit{structural features} are size, depth and
breadth.
Size counts the number of nodes, either the nodes representing distinct, or the nodes representing non distinct entities. In the case
of tree-structured networks, depth is the maximum distance from the root to any of the leaves,
while depth counts the number of nodes at a level of the tree.
There are also features of individual nodes, such as the centrality of nodes, as measured by the  degree centrality, closeness centrality and betweeness centrality.
Distance-based features consider the shortest path distances between pairs of nodes.
One such feature is structural virality defined as the average Wiener index, i.e., the average shortest path distance between all pair of nodes \cite{Goel16}. Structural virality tends to be maximized for large networks that have become that way through
many small branching events over many generations.
Other structural features are about the local structure of the graphs, captured for example using modularity.

\textit{Temporal features} have also been considered.
In this case, networks are annotated with time information, for example, in a cascade network, each node is annotated with the time when the corresponding item was posted.
Common temporal features are lifespan, speed, and heat.
Lifespan captures the duration of spreading a news item and can be measured as the difference between the time associated with the first and the last node. Speed refers to how fast information is spread, and may be measured as the time needed to reach a specific number of users, a specific level of the tree, or the value of some other structural feature.
Heat refers to the number of nodes, e.g., comments, that appear within a specific time interval.

Several other types of features have also been studied, besides structural and temporal ones. A family of such features looks at the users involved in the propagation and computes various statistics about the distribution of their features in the network, such as the percentage of bots. A fairly recent work studies the habitual nature of misinformation propagation. Specifically, Facebook users were found to automatically react to the familiar platform cues, sharing misinformation out of habit \cite{ceylan2023sharing}. For example, headlines were presented in the standard manner (i.e., Facebook format of a photograph, source, headline) with the sharing response arrow underneath, and users clicked the shared button habitually.
Analogously the distribution of content features, such as linguistic aspects and sentiment, have also been exploited \cite{Shu20}. Finally, there are features about the relationships between the different networks involved.
For example, one such feature is whether is more likelihood of retweeting is larger between friends.
%Table \ref{tab:prop-features} summarized the different types of features.

\begin{footnotesize}
\begin{table}
\caption{Propagation network features}
\label{tab:prop-features}
\setlength{\tabcolsep}{3pt}
\begin{tabular}{|l|p{310pt}|}
	\hline
   Name &  Description  \\\hline
     \multicolumn{2}{|c|}{Structural Features} \\
    \hline
	Size &  Number of nodes, or unique nodes  \\
	Depth &  Number of edges from the root to a leaf node   \\
	Breadth  &  Number of nodes at some tree level  \\
	Node centrality & Node degree, betweeness  \\
	Virality & Average distance between all pairs of nodes  \\
	 & (aka average Wiener index)  \\
	Modularity & Clustering
 \\ \hline
	  \multicolumn{2}{|c|}{Temporal Features} \\ \hline
	Duration &  The longest interval  \\
Heat  &  Number of users posting at a specific time interval \\
	Speed  & The time it takes to reach a specific depth, number of nodes, etc
	\\ \hline
	\multicolumn{2}{|c|}{Other Features} \\ \hline
	User feature distribution & Distribution of features of the users involved in the propagation  \\
        User habitual reactions & Consistent form and interaction widgets of a platform for all news trigger habitual reactions from users \\
Content (item) feature distribution & Distribution of features of the items involved in the propagation   \\
	More than one network & Correlations/relationships between different networks  \\
	 \hline
\end{tabular}
\end{table}
\end{footnotesize}

Various studies highlight the difference in structural, temporal and other features between graphs related to fake and true news.
The authors of \cite{vosoughi2018spread},
used the retweet network of all fact-checked cascades
on Twitter from  2006 until 2017. Their study showed that fake news was propagated six times faster than real news.
Specifically, it was shown that it
took the truth approximately six times as long as
falsehood to reach 1,500 people and 20 times as
long as falsehood to reach a depth of ten.
Falsehood was also diffused significantly
more broadly and was retweeted by more unique
users than the truth at every depth.
False political news diffused deeper more quickly, and reached
more people faster.
In addition, the inclusion of bots in social media accelerated the
spread of both true and false news. However, it affected their
spread roughly equally.

% \cite{Safarnejad20} is [170] from medical survey
In the medical domain, work in \cite{Safarnejad20} constructed cascade
networks of retweets during the Zika epidemic in 2016. They report statistically significant differences in 9 network metrics between real and fake groups of cascade networks.
Fake news cascade networks were in general more sophisticated than those of the real news.
The metrics they consider are the number of nodes,  number of distinct nodes,  diameter, Wiener index, structural virality, out-degree distribution, the largest betweenness centrality, modularity, the relation between follower networks and retweet, i.e., the difference in probability of retweet between followers and non-followers.

In a recent study, the authors of  \cite{di2022health} examine and analyze several distinct groups of features (introduced in \cite{Shu20}) and various machine learning techniques to assess misinformation in online medicine-related content. Amongst others, they are using propagation-network features. They have divided these features into four categories: Structural features, temporal features, linguistic features, and engagement features. Structural features capture the network structure of users and can be depth, breadth, or out-degree. Temporal features include the duration of dissemination, the average speed of dissemination, and the average speed of response. Linguistic features are variables that capture the linguistic aspects of messages that interact with information dissemination.
% To assess sentiment in these texts they use VADER\footnote{\url{https://github.com/cjhutto/vaderSentiment}} which is a lexicon and rule-based sentiment analysis tool for social media content.
Finally, engagement features are variables that assess the level of appreciation. After the feature construction, they use binary classification to distinguish health information from misinformation. In their results, they report that the random forest classifier had the best performance among the rest.
% check whether to add more here +

Some propagation-based methods use trees or tree-like structures to capture the propagation of an article on social media. For instance, in \cite{wu2015false} authors propose a graph-kernel-based hybrid SVM classifier that captures the high-order propagation patterns. Specifically, they model the message pattern propagation as a tree that reflects the relation between reposts and their authors. Moreover, they propose a random walk graph kernel to model the similarity of propagation trees. They combined the graph kernel with a radial basis function kernel to build a hybrid SVM classifier. The structural features used for classification were features such as the total number of nodes, and the maximum or average degree of the graph. They achieved significantly better classification accuracy than other models for the early detection of false rumors.

\subsection{Beyond propagation features}
Recent research leverages complex graphs beyond the propagation networks to leverage rich contextual information about
the news and users involved in spreading misinformation. Entities involved in the news propagation include the original news item, the related posts, publishers and users. Edges
in the graphs correspond to the various relations between these entities.
Initial feature representations of the nodes of the graphs are created using content and other features.
These representations are then enhanced using the relations in the graphs.
Approaches differ in the types of entities, relations, and  graphs as well as in the methods used to create latent representations.
Below we present some representative approaches that use such complex networks for fake news detection \cite{Cui22,Dhanasekaran21,min22,Paraschiv22}.

%[50] from medical surevy
Hetero-SCAN \cite{Cui22} builds a heterogeneous graph capturing various types of nodes and interactions. Specifically, the graph consists of three
types of nodes: publishers who create the news, users who tweet the news and
news. There are four types of edges: citation edges between publishers, publish edges between publishers and news,
tweet edges between users and news, and friendship edges between users.
The graph is used to learn feature representations for the news nodes which are used as input to a fake news classifier.
Initial feature representations are constructed for the three types of nodes using  appropriate text and graph embedding methods.
Then, the feature representations for the news nodes are enriched using instances of two types of meta-paths: news $\rightarrow$  publishers $\rightarrow$ news and news $\rightarrow$  users $\rightarrow$ news.
KG-based embeddings are used, in particular TransE, refined for this task.
The premise behind  using these specific meta-paths is that the
validity of a specific news items is related to the authenticity of both its publishers and its users as well as the news that the
publishers and users respectively publish and tweet.

% [59] from medical survey
The SOMPS-Net framework  builds an end-to-end binary fake news classifier \cite{Dhanasekaran21}.
The entities are news articles, tweets and retweets of the articles, users, and publishers.
SOMPS-Net consists of two components, the social interaction graph component (SIG) and the publisher and news statistics component (PNS).
The SIG component has several subcomponents.
Initial textual embeddings of each article and its related tweets and retweets are generated using  GloVe.
Feature vectors are constructed for the users based on information from their profile and history,
such as the number of their tweets and friends.
For each article, two social connectivity graphs are built, one with the users that tweeted about the article and one
with the users that re-tweeted about it. There is an edge from a user $u$ to a user $v$, if $u$ follows $v$ weighted
by the number of common followers and followees of $u$ and $v$.
A graph convolution layer is applied on the two graphs to obtain graph embeddings.
A cross attention component is then applied to capture the correlation between the news embeddings and the graph embeddings.
The PNS component includes statistical features about the article, such as the total number of tweets, and
some features of its publisher. The PNS feature vector is fused with the output of the SIG component and the result is used
to predict the validity of a news item.

% [144] from medical survey
Fake news detection  is formulated as a binary graph classification problem in the Post-User Interaction Network
(PSIN) model \cite{min22}.
The entities are news items, posts, and users.
An interesting aspect of the approach is that they  adopt an adversarial
topic discriminator to
improve the generalizability of the model to emerging topics.
A heterogeneous directed graph is built for each news items.
The graph includes two types of nodes: the users and the posts involved in the dissemination of the item.
There are three types of edges: propagation edges between posts, follower edges between users, and post edges from
users to posts.
The heterogeneous graph along with the topic of a news item are given as input to a classifier that determines whether the
item is fake or not.
To process the graph, the graph is decomposed into three parts: post
propagation tree, user social graph and post-user interaction graph.
Node representations are build using both textual features and meta features (e.g., number of likes).
Each part of the graph is processed individually,
using appropriate graph attention (GAT) models. At the end, the concatenated and pooled representations
is  fed into a veracity classifier and an adversarial topic classifier.

% [156] from medical survey
A different approach is taken in \cite{Paraschiv22} where a  new graph data structure, termed meta-graph is proposed.
The nodes of the meta-graph correspond to retweet cascades. Each cascade disseminates a root tweet along with the web and/or media URL(s) embedded in that tweet.
There is an edge between two nodes (i.e., cascades), if there are common users in the two cascades.
%Edges are weighted by the number of common users.
The node features contain information about the cascade graph structure, obtained by applying a graph embedding algorithm to each individual cascade, textual information potentially including topic and sentiment, profile user information, and relational user information, extracted from the social network.
The edge features encode relationships between cascades, such as structural similarity,
number of common users, or content similarity.
Two formulations of the problem of fake news detection are proposed.
The first formulation is as a graph classification problem where each graph corresponds to a single cascade, i.e, the meta-graph is not used.
The second formulation is as a node classification
problem at the meta-graph level.
The meta-graph node classification approach slightly outperforms the graph classification approach in accuracy.
Various GNN-based classifiers are used in both approaches.

Complex networks are used to determine whether an item refers to a known
misconception  in \cite{weinzierl2021automatic}. Specifically,  given a set of misinformation targets (MisTs), each addressing
a known misconception about the COVID-19 vaccines, the authors
address the problem of  discovering
automatically which tweets contain misinformation and the one or more MisT
they refer to.
They construct a  Misinformation Knowledge Graph (MKG) whose nodes are tweets that contain misinformation.
An edge between two tweets indicates that they share the same misconception, and the tweets are
labeled with the corresponding MisT.
For each MisT, a separate, fully-connected graph (FCG) is generated
spanning all tweets that refer to this MisT.
Then the problem of discovering whether an unconnected tweet contains misinformation is cast as a link prediction problem: predict whether
there is a link  from the tweet to any FCG.
Representations of the tweets are learned using several KG embedding models.
To capture linguistic content,  embeddings of MisTs and tweets are also created using COVID-Twitter-BERT-v2 and projected
in the embedding space of the MKG.
The link
ranking functions from each KG embedding
model are used for predicting a link between any tweet that may contain
misinformation and tweets that share a known MisT.

Finally, an additional type of information that can be exploited is the fact that an item may take a  different stance on a topic.
To exploit this information, some approaches introduce stance networks where different types of edges are used to exploit agreement or disagreement between nodes (e.g., news items, or users).
For example, the approach proposed in \cite{jin2016news} exploits a signed network of tweets to predict
the credibility of a tweet. Credibility  is a numeric value in $[-1, 1]$ with positive values
corresponding to valid information and negative values to fake one. There are two types of edges:
positive edges between tweets taking the same viewpoint and negative edges between
tweets taking different viewpoints.
Initial credibility values are assigned to some tweets. Then, credibility propagation is formulated
as a graph optimization problem. The solution is based on the assumption that
tweets with similar viewpoints should have similar credibility
values while tweets with opposing viewpoints should have
opposite credibility values.

Table \ref{tab:prop-approaches} summarizes the approaches that exploit the network information.

\begin{footnotesize}
\begin{table}
	\caption{Propagation-based Approaches.}
	\label{tab:prop-approaches}
	\setlength{\tabcolsep}{3pt}
	\begin{tabular}{|p{80pt}|p{150pt}|}
		\hline
		Problem &  Proposed Approach   \\\hline
		Fake news detection & Explicit Features + Classification \cite{di2022health}, \cite{wu2015false} \\
		 & Embeddings + Classification \cite{Cui22},\cite{Dhanasekaran21},\cite{min22},\cite{Paraschiv22} \\
		 & Embeddings + Node Classification (labeling) \cite{Paraschiv22} \\\hline
		Fake item detection &  Link prediction \cite{weinzierl2021automatic} \\
			 & Graph optimization \cite{jin2016news} \\ \hline
	\end{tabular}
\end{table}
\end{footnotesize}

\section{Source-based Methods}
\label{sec:sources}

\begin{footnotesize}
\begin{table*}[h!]
\centering
\caption{Features used in source-based approaches}
\begin{tabular}{|l | p{310pt} | l |}
 \hline
 Feature Type & Description & Citation \\
 \hline\hline
 \multirow{2}{*}{User} & User Type, Gender, Age, Location, Average number of posts, Post frequency, Repost frequency, User activity
      &  \cite{di2022health,kar2021no,DBLP:journals/corr/abs-1809-00557, heidari2021bert},\\
      & Number of days since the account was created, Account verification status, Account protection status,  & \cite{Nastasi2018BreastCS}\\

\hline
   \multirow{3}{*}{Content} & Biomedical terms inside users descriptions, Number of characters in users name/ description,
   &  \cite{kar2021no} \\
   & Number of URLs in users description, Existence of official URL,Number of likes, Number of likes received & \\
   &readability, number of adverbs and numbers &\\
\hline
 \multirow{2}{*}{Network}  & Number of  followers, followees & \cite{di2022health,heidari2021bert,kar2021no}\\
    & Average numbers of biomedical terms inside followers and followees description  & \\
 \hline
  \multirow{1}{*}{Behavioral}  & Discussion initiation, Interaction engagement, Influential scope, & \cite{Zhao2021DetectingHM}\\
    &  Relational mediation, Informational Independence &\\
    \hline
  \multirow{1}{*}{Temporal}  & Interval entropy between user's posts & \cite{DBLP:journals/corr/abs-1809-00557}\\
  \hline
\end{tabular}
\label{table:source-based features}
\end{table*}
\end{footnotesize}

Most fake news detection algorithms focus on finding clues from the content of news, which is not always effective since fake news are intentionally written to mislead users by imitating valid news. Therefore, it is helpful to explore ancillary information to improve detection. One such piece of information is the credibility of the news sources. Source-based fake news detection uses methods to assess the credibility of the source of news.
%The source can refer to sources that create the news (authors), sources that publish the news (publishers), and users that spread the news on social media (social media users).
The source can refer to publishers, such as news portals that publish the news, and the social media users who either create new content or post existing news articles published by the publishers.

Thus, we divide source-based methods into two main categories, based on the type of source: Methods that focus on the credibility of publishers, and methods that focus on the credibility of social media users \cite{zhou2020survey}. The latter considers also the case of social bots and trolls.

%relationships between publishers and authors in order to assess source credibility, and the ones which focus on the credibility assessment of social media users \cite{zhou2020survey}.

%We will consider the following source based methods: Methods that assess the credibility of publishers, methods that

\subsection{Publisher Credibility}

Assessing the credibility of publishers is closely related to the problem of assessing the credibility of web domains. This is a problem that has been considered in the literature for ranking web pages \cite{pagerank,hits} and fighting spam (e.g., see \cite{ntoulas2006detecting,liu2015towards}). We aim to assess the credibility with respect to the truthfulness of the information.

%To assess the credibility of web domains, several approaches are utilized, that aim to identify characteristics and patterns that are indicative of fake news.
Assessing the credibility of web domains relies on identifying characteristics and patterns that are associated with fake news. A variety of approaches has been considered,
such as content analysis, link analysis, natural language processing, and machine learning algorithms.
%These methods help identify characteristics and patterns that are often associated with spam.
We distinguish two main categories of methods:
%used to identify websites that contain fake news:
methods that use content-based features, and methods that use graph-based features.

Content-based features extract specific information from the text of websites to distinguish between valid and non-valid websites. For example, in \cite{baly2018predicting} to assess the credibility of a source they extract features from a sample of the website articles, the corresponding Twitter account, the Wikipedia page, and the web traffic information to the domain. The study found that the Twitter account had the most impactful features.

Graph-based features are used to analyze the structure of the network of Web pages and the links between them. This can involve identifying patterns of link exchanges and websites that are highly connected to other websites that include fake news. Consequently, most graph-based fake Web site detection algorithms follow the Pagerank intuition by iteratively propagating scores through the links in the Web graph.
%traversing links in the Web graph while constantly updating each node’s score.

Using the graph of web pages and links, it is also possible to detect communities of websites that are interconnected and may be involved in spreading fake news. \cite{chen2020proactive} focuses on detecting fake news on web pages utilizing both graph and content-based features. They associate each domain with a group of Twitter users who tweet about it and create a graph by linking domains whose user sets have a Jaccard similarity above a specified threshold. Then a classifier is used with three types of features: markup, readability, and morphological. The top features associated with real news include advanced HTML tags and high readability scores, while the top features for fake news include italic fonts and underscore.

%\subsubsection{Relations between authors or publishers}
Several studies have indicated that there is a tendency for news publishers (and journalists) to form homogeneous networks. These networks can be analyzed to identify patterns and assess the credibility of unknown authors or publishers. If an unknown author or publisher is found to be connected to a network of individuals that have spread misinformation, it may be reasonable to question their credibility \cite{zhou2020survey}.

In \cite{horne2019different} authors explore how the credibility of the news publishers relates to the communities they form.
They collected publisher domains and articles from both mainstream outlets and alternative sources. They created a content-sharing directed network, where nodes are news sources and edges are articles copied between the sources. Edges are weighted by the number of articles copied and they are directed from the original source to the source that copied. To identify the communities in the network, they used modularity maximization algorithm \cite{leicht2008community}. The credibility of the news sources was then evaluated by using several tools, News Guard, Media Fact Check, Allsides and BuzzFeed. They combined the outcomes from these tools to generate a score ranging from -1 to 1. Sources with score less than -0.6 were labeled as Non Credible, those with score more than 0.6 labeled as Credible, while sources in between were labeled as Unknown.
%By utilizing the network,
They then studied the relationship between the community structure in the network and the credibility of sources, and they identified four main behaviors: echo chambers, context mixing, competing narratives, and counter-narratives. They concluded that news publishers face challenges in maintaining their credibility. Alternative news sources tend to promote conflicting narratives, eroding trust in mainstream media. Some right-wing sources mix mainstream and conspiracy content, potentially lending false credibility to unreliable sources. News is homogeneous within communities, creating different spirals of sameness and amplifying alternative narratives.
%, when formulated as a network, news producers
%news publishers share content in tightly connected communities %which represent distinct parts of the media ecosystem.

\subsection{Social media users}
%Source based methods that focus on the credibility assessment of social media users rely on the idea that users with low credibility tend to spread fake news. Users in social media can be divided into humans (non-automated users) and bots (fully automated accounts).

We now turn our attention to the credibility assessment of users in social media. We divide social media users into two categories: humans (normal, non-automated users) and bots (fully automated accounts).

The assessment of social media user credibility is based on a diverse range of source-based features, including \emph{user, content, network, behavioral,} and \emph{temporal} characteristics. User features include basic information such as age, profile, and interests, while content features pertain to the textual information found in the user profile, posts, and comments. Behavioral features refer to the user behavior on the platform, such as their discussion initiation, interaction engagement, influential scope, and more. These features can provide insights into the user's level of engagement with the platform and their role in the community.
Timing features, such as interval entropy, can also be analyzed to determine if the user is sharing content in a manner consistent with typical human behavior.
%Timing features, such as interval entropy, can also be analyzed to determine if the user is sharing content in a manner consistent with typical human behavior.
Table \ref{table:source-based features} presents a summary of the most commonly used features in source-based methods.

In \cite{kar2021no} authors propose an approach for the early detection of fake news about COVID-19 from social media, such as tweets for multiple Indic-Languages besides English, using both \emph{content} and \emph{source} based features. To determine the credibility of the source they use \emph{user features} from Twitter and \emph{network features}. Some of the features as reported in  Table \ref{table:source-based features} are the number of characters and the total number of URLs in the description of the user, whether the user has an official URL or not, the number of followers, the number of days since the user's latest tweet, whether the user account is verified officially, etc. After experiments on various combinations of features and classifiers, they concluded that user features are effective at determining the credibility of the news that the user spreads.

Another approach \cite{Zhao2021DetectingHM} focuses on \emph{behavioral features} of users in order to detect misinformation spreading on Zibizheng Ba, a Chinese forum about autism on Baidu Rieba. Specifically, they measured discussion initiation (the tendency of a user to initiate a new discussion), interaction engagement (how much a user is interacting with other users), influential scope (the communication ability of a user) that is measured by using the degree centrality in the social network of the user, relation mediation (ability of control of communication in the community), measured by betweenness centrality and informational independence (ability of user to instantly communicate with others without going through many intermediaries) by measuring closeness centrality.

Some other medical fake news detection approaches that make use of source-based methods examine the characteristics of users who are sharing unverified cures for various illnesses. In \cite{DBLP:journals/corr/abs-1809-00557} authors are focusing on misinformation spread by users about cancer treatments on Twitter. They separate users into two groups, the \emph{rumor} group with users that have posted some of the "treatments" and the \emph{control} group who have posted generally about cancer but none of these topics. They measure and compare \emph{user features}, \emph{content features}, and \emph{temporal features}. They observed that the most important and discriminative features of content-based features are readability, number of adverbs and numbers, user verification regarding the user features, and interval entropy regarding temporal features. Also, they concluded that most viral rumors were on "cures" that involved food or drink.

The importance of the knowledge background of a user when dealing with healthcare information dissemination is studied in \cite{Nastasi2018BreastCS}. The authors analyze Twitter users and their content in the dissemination of medical news around mammography, and the ensuing discussions.
%and the quality of evidence around discussions.
They separate users in non-healthcare, health organizations, commercial healthcare organizations, physicians, non-provider health professionals, media organizations, health advocates, cancer specialists, and non-physician providers. They use \emph{user features} like gender, age, location, and \emph{content features} to analyze and determine who is sharing information about medical topics that may not be valid or verified. They conclude that non-healthcare users are the primary drivers of such conversations and tend to make claims without providing any references.

\subsection{Trolls}
A \emph{troll} is an individual with antisocial behavior who intentionally creates posts and comments in social media to provoke emotional reactions from others. Such individuals often use aggressive or offensive language in order to slow down the natural progression of the conversation. The purpose of a troll is to offend people, dominate discussions and manipulate opinions of people \cite{Tomaiuolo2020ASO}.

%Trolling has three main components: a pseudo-intention which is the surface level intention of a troll's behavior, a real intention which is to provoke and disrupt conversations and a stimulus which is the trigger that prompts a troll to engage, such as a particular topic or a person \cite{Tomaiuolo2020ASO}. Unfortunately, some users are unable to distinguish a troll's true intention and may become emotionally involved, sharing private information or even engaging in criminal acts like cyberbullying which can lead to suicide.

The prevalence of trolls has grown in recent years, causing concern for both individuals and society. Identifying trolls is a challenging task, as trolls go to great lengths to conceal their true identity and their actions. Many online communities such as Reddit have developed strategies to identify and remove harmful content like moderation teams that monitor and delete inappropriate messages manually \cite{Tomaiuolo2020ASO}, \cite{FORNACCIARI2018258}.
%Another concentrated effort for identifying trolls was conducted by the US Congress
US Congress also published a catalog of 2,752 Twitter accounts which were allegedly linked with the Internet Research Agency, a Russian-based organization referred as a "troll farm" \cite{DBLP:journals/corr/abs-2001-10570}, \cite{RUFFO2023100531}, in an effort to deter what was perceived as a systematic campaign for mass manipulation of public opinion by Russia.

%One of the examples was from Russia that has been accused by the US Congress of conducting a systematic mass manipulation of public opinion using both trolls and bots. Along with the accusation a catalog of 2,752 Twitter accounts found which claimed to be linked with the Internet Research Agency, a Russian-based organization referred as a "troll farm" \cite{DBLP:journals/corr/abs-2001-10570}, \cite{RUFFO2023100531}.

%One of the defining characteristics of trolls is their anonymity. Many trolls use pseudonyms or operate from anonymous accounts and this can make it difficult to identify them. Many online communities such as Reddit have developed strategies to identify and remove harmful content like moderation teams that monitor and delete inappropriate messages manually \cite{Tomaiuolo2020ASO}, \cite{FORNACCIARI2018258}.

%The prevalence of trolls has grown in recent years, causing concern for both individuals and society. Various research have been conducted to detect and identify trolls with automatic methods \cite{DBLP:journals/corr/abs-2112-00443}, \cite{DBLP:journals/corr/abs-2001-10570}, \cite{FORNACCIARI2018258}, \cite{DBLP:journals/corr/abs-2103-09054}, \cite{Engelin2016TrollD}.

There has been considerable research effort on automatic troll detection approaches \cite{DBLP:journals/corr/abs-2112-00443}, \cite{DBLP:journals/corr/abs-2001-10570}, \cite{FORNACCIARI2018258}, \cite{DBLP:journals/corr/abs-2103-09054}, \cite{Engelin2016TrollD}.
Troll detection involves several methods such as analyzing the language used in posts or comments using NLP techniques to determine the sentiment and emotional tone of messages. User behavior analysis is another method  that can be used as trolls often exhibit certain patterns of behavior, such as continuously posting negative comments or targeting specific users. Network analysis is also used as trolls often operate in groups or networks, and by analyzing social network connections, it is possible to identify groups of users engaging in trolling behavior \cite{Tomaiuolo2020ASO}.

For instance in \cite{DBLP:journals/corr/abs-2112-00443} authors train a model on 335 Russian-sponsored troll accounts identified by Reddit. They demonstrated that troll accounts exhibit specific interaction patterns, such as a higher frequency of replies among each other, and submissions with the same title. They considered accounts from the top subreddits with trolls (r/News, r/Politics, r/Bitcoin). They concluded that the most important features include the number of submissions, account age, the fraction of submissions with the same title as known troll submissions, the fraction of comments that reply to a troll account, etc.

In \cite{FORNACCIARI2018258}, the TrollPacifier tool is proposed for detecting trolls. Thus model was trained on Twitter data using six groups of features: sentiment analysis, time and frequency of actions, text content and style, user behaviors, interactions with the community, and advertisement of external content. Community and advertisement group of features were found to outperform the rest. Additionally, in \cite{DBLP:journals/corr/abs-2103-09054} the authors focus on troll detection in comments via sentiment analysis and user activity data on the Sina Weibo Chinese platform.

Trolls can also play a significant role in the spread of health misinformation in social media and lead to confusion of the public. For example, during the COVID-19 pandemic, trolls were very active in spreading health misinformation on a wide range of topics, such as treatments, vaccines and the origins of the virus \cite{10.1145/3442167.3442169}. Due to the serious consequences of this kind of misinformation, several troll detection tools have focused on the health domain. For example, in \cite{10.1145/3442167.3442169} the authors collect a variety of Twitter content features to detect whether a tweet was meant to troll. As features, they use trolling hashtags and specific
strings that were commonly used to advance trolling topics, sentiment analysis features, and user behavioral features.

\subsection{Social Bots}
\iffalse%%%%%%%%%%%%%%%%%%%%%%%%%
We should probably also include trolls.
OUtline:
What are social bots. Positive uses of social bots. General remarks on negative uses of Social Bots. Their role in fake news propagation (difference of opinions). The problem of bot detection.
Pandemic and health issues with bots. Health-specific issues, studies.
    Trolls: Different from bots but related. The problem of troll detection. The use of trolls for health issues.
\fi%%%%%%%%%%%%%%%%%%%%%

A \emph{bot} is a software program designed to automate tasks or simulate human behavior on the internet. There is a wide range of different types of bots, such as spiders or crawlers, shopbots, knowbots, chatbots and social bots.

A social bot is a type of bot that imitates human behavior in social media platforms. Social bots can be utilized for both positive and negative purposes \cite{DBLP:journals/corr/abs-2004-09531}. Some of the positive purposes are disseminating accurate and helpful information to a wide audience, promoting products or services in a cost-effective manner and providing customer support \cite{DBLP:journals/corr/abs-1710-04044}. Also, social bots have helped in emergency management like in publishing earthquake warnings in Japan \cite{hofeditz2019meaningful}.

However, they have also been used for negative purposes such as spreading misinformation and propaganda to manipulate public opinion, scamming people and creating fake accounts to manipulate or increase follower counts. Social bots, in some cases have some similarities with trolls. Both bots and trolls can be used to spread fake news and propaganda and influence public opinion.
They are often designed to spread malicious software, create confusion and distraction, and manipulate social media discourse with rumors \cite{khaund2018analyzing}. They can infiltrate into a group of people and change their view of reality \cite{10.1145/2818717}. Some notable examples are the role of social bots on the Brexit debate where they influenced users opinion about the election \cite{DBLP:journals/corr/abs-1710-07562} or the effect of social bots on Twitter during the 2016 U.S. election. Some studies claim that 64\% to 79\% of Donald Trump's followers were fake. Such accounts made opinions appear more widespread \cite{8537833}.

The effect of bots on health-related news can be really harmful. Bots can spread fake or misleading information about health topics and cures which can lead to confusion and harm public health. During the COVID-19 pandemic, the majority of social bots have been tweeting about COVID-19 \cite{info:doi/10.2196/26933}. However, there is limited evidence of regarding their stance on these topics. Some bots were spreading accurate up-to-date information about the pandemic and others were spreading misinformation, conspiracy theories, fake treatments and rumors about the vaccine. It is shown that the activity level of social bots was consistent with that of human users \cite{ijerph19031651}. Nevertheless, in \cite{article} authors by utilizing a combination of machine learning and network science methods on a collection of tweets about the pandemic in the Philippines and the United States, they  indicate a significant association between bot activity and increased levels of hate speech in both countries, particularly within densely populated and socially isolated communities. Additionally, they highlight the significance of adopting a global perspective in computational social science, especially in the midst of a worldwide pandemic and infodemic, which have universal yet unequal impacts on societies.

To detect bots, various techniques, similar to those used in troll detection, are employed \cite{info:doi/10.2196/26933}. Profile metadata, such as account age and username, along with user activity, temporal patterns, and posts content. Bots tend to retweet more frequently and have shorter time intervals between tweets. Additionally, network patterns, such as friend/follower connections, hashtag use, retweets, and mentions, are used to identify social bots that exploit network homophily. Bot detection is getting harder as bots evolve and make it more difficult to distinguish them from human behavior \cite{10.1145/2818717}.

In \cite{egli2021bad}, the authors collected tweets and analyzed user activity by measuring the number of tweets, retweets and replies. They also examined the types of software used to generate tweets (some users use Twitter clients, some others use cross-posting through other platforms, while commercial actors use marketing automation software). Finally, they developed a classifier that can identify social bots based on the user activity, which is measured as the sum of tweets, retweets, and replies posted by a user and visibility which is measured as the number of retweets and replies received by a user.

In \cite{heidari2021bert} the authors used two datasets, the COVID-19 dataset \cite{cui2020coaid}, and a dataset with human accounts and three types of bots including social spam bot, fake followers, and traditional Twitter spambots. This dataset consists of user features such as follower-count, retweet-count, reply-count, number of shared URL etc.
%The author used bidirectional encoder representations from transformers (BERT) in order
They used BERT to classify the origin posts into bot or not bot. They concluded that both bot and human accounts have very similar behavior in spreading fake or real news. Also, they observed that bots are not the primary source of fake news in the COVID-19 pandemic.

%[Copied from Misinformation, manipulation, and abuse on social media
%in the era of COVID‑19 https://link.springer.com/content/pdf/10.1007/s42001-020-00094-5.pdf?pdf=button]
%Yet, another aspect of online manipulation—that is, automation and social bot
%interference—is tackled by Uyheng and Carley in their article “Bots and online hate
%during the COVID-19 pandemic: Case studies in the United States and the Philippines” (Uyheng, J., \& Carley, K. M. (2020). Bots and online hate during the COVID-19 pandemic: Case
%studies in the United States and the Philippines. Journal of Computational Social Science.). Using a combination of machine learning and network science, the
%authors investigate the interplay between the use of social media automation and the
%spread of hateful messages. They fnd that the use of social bots yields more results
%when targeting dense and isolated communities. While the majority of extant literature frames hate speech as a linguistic phenomenon and, similarly, social bots as an
%algorithmic one, Uyheng and Carley adopt a more holistic approach by proposing a
%unified framework that accounts for disinformation, automation, and hate speech as
%interlinked processes, generating insights by examining their interplay. The study
%also reflects on the value of taking a global approach to computational social science, particularly in the context of a worldwide pandemic and infodemic, with its
%universal yet also distinct and unequal impacts on societies.

\section{Datasets and Tools}
\label{sec:DET}
In this section, we discuss publicly available datasets for fact-checking and fake news detection for the medical domain, along with well-known fact-checking services and tools.
\subsection{Datasets}
\label{sec:Datasets}

Table \ref{tab4} lists 24 health-related fake news and fact-checking datasets that are publicly available. Most of these datasets have been constructed from Twitter posts and user interactions, that exploit also online news articles. The only exceptions are \cite{heley2022missing} and \cite{wadhwa2022redhot} which contain a collection of claims constructed from Reddit r/coronavirus and submissions from various health-related subreddits correspondingly.

A crucial step in annotated datasets is how the annotation is conducted. There are two basic approaches to annotation, manual and automatic. In manual annotation, the annotation can be done by experts of the domain \cite{memon2020characterizing, ceron2021fake, ginsberg2021report},  which is an expensive and time-consuming process, by crowd-sourcing techniques like in \cite{heley2022missing} or the highly moderated Reddit subreddits such as  r/coronavirus that require sources of claims, and finally by the authors themselves using fact-checking services or a search engine \cite{kar2021no, waszak2018spread, zhou2020recovery}. Regarding automatic approaches, some datasets have been annotated by using the inference capabilities of already trained models \cite{wuhrl2021claim}. Other datasets have used DL retrieval models to manually annotate only a subset of the dataset, which includes the most relevant tweets to a misinformation target \cite{weinzierl2022vaccinelies}. Finally, \cite{abdul2020mega} uses distant supervision, where a database or a KG is used to collect examples for the relations to be extracted. Then based on these examples and whether they coexist in a sentence, the user post or news article can be annotated accordingly.

%Table \ref{tab4} also reports the best F1 score in the surveyed literature for each dataset.
The performance of the various models and approaches proposed in the literature over these datasets spans across a variety of scores. For example, the F1 scores can range between 33\% - 99\%, reported in \cite{dharawat2022drink} and \cite{cui2020coaid} respectively. This showcases the importance of the intrinsic properties of the datasets used for evaluation purposes (e.g., what sources they use, their size, how the dataset was labeled, etc.), and the task at hand (binary or more refined classification like multi-class and multi-label classification).

Finally, most of the approaches introduce their own datasets, making the comparison between the various approaches difficult and inconsistent. As a result, the community involved in the domain of medical fake news detection needs a dataset that could be used as a benchmark for evaluation purposes.

%The next section offers a brief discussion of some common evaluation metrics used in classification tasks, including f1 scores.

\begin{footnotesize}
%\onecolumn
\begin{longtable}[]{|p{40pt}|p{75pt}|p{90pt}|p{50pt}|p{155pt}|}
\caption{Public datasets related to medical fact-checking, fake news detection, and social media}
\label{tab4}\\
\hline
\textbf{Work}
    &  \textbf{Type}
    & \textbf{URL}
    & \textbf{Annotation}
    & \textbf{Characteristics}
%    & \textbf{Used}
%    & \textbf{F1 score}
\\\hline\hline
% 1st
BioClaim \cite{wuhrl2021claim}
    & -Biomedical claims
        \newline -Tweets
    & \scriptsize{\url{https://www.ims.uni-stuttgart.de/documents/ressourcen/korpora/bioclaim/WuehrlKlinger-Bioclaim2021.zip}}
    & Manual
    & 1200 tweets for implicit and explicit biomedical claims (the latter also with span annotations for the claim phrase), including COVID-19, measles, cystic fibrosis, and depression
%    & -
%    & 90\% \ci
\\\hline
% 2nd
NewsClaims \cite{reddy2022newsclaims}
    & -COVID-19 claims
        \newline -News
    & \scriptsize{\url{https://github.com/blender-nlp/NewsClaims}}
    & Manual
    &889 claims annotated over 143 news, including claim sentences, claim objects, claim stances, claim span, and claimer detection
%    & -
%    & 90\% \cite{li2020mm}
\\\hline
% 3rd
RedHOT  \cite{wadhwa2022redhot}
    & -health issue claims
        \newline -Reddit
    & \scriptsize{\url{https://github.com/sominw/redhot}}
    & \revision{Manual (Crowd-sourced \& Experts)}
    &22,000 posts from 24 medical condition subreddits, annotated as claims, personal experiences, and questions
%    & -
%    & 45.16\% for claims \cite{wadhwa2022redhot}
\\\hline
% 4th
PUBHealth  \cite{kotonya2020explainable}
    & -health issue claims
        \newline -claims from fact-checking sites and news
    & \scriptsize{\url{https://github.com/neemakot/Health-Fact-Checking}}
    & Manual \newline Automatic
    &27,578 fact-checked claims, 9,023 claims from Associated Press and Reuters News, and 2,700 claims from Health News Review. Labels are true, false, mixture, and unproven
%    & -
%    & 70.52\% for claims \cite{kotonya2020explainable}
\\\hline
% 5th
HealthVer  \cite{sarrouti2021evidence}
    & -health issue claims
        \newline -search engines snippets
    & \scriptsize{\url{https://github.com/sarrouti/HealthVer}}
    & \revision{Manual (Authors)}
    &14,330 evidence claim pairs, labeled as support, refute, and neutral
%    & -
%    & 79.60\% for claims \cite{sarrouti2021evidence}
\\\hline
% 6th
ANTi-Vax \cite{hayawi2022anti}
    & -COVID-19 vaccine misinformation
        \newline -Tweets
    & \scriptsize{\url{https://github.com/SakibShahriar95/ANTiVax}}
    & \revision{Manual (Experts)}
    & 15 million COVID-19 vaccine-related tweets and 15 Thousand labeled tweets for vaccine misinformation detection (binary classification)
%    &
%    & 98\% \cite{hayawi2022anti}
\\\hline
% 7th
CMU-MisCOV19 \cite{memon2020characterizing}
    & -COVID-19 vaccine misinformation
        \newline -Tweets
    & \scriptsize{\url{https://zenodo.org/record/4024154/files/CMU_MisCov19_dataset.csv.zip}}
    & \revision{Manual (Authors)}
    & 4573 annotated tweets, comprising of 3629 users. Labels are: Irrelevant, Conspiracy, True Treatment, True Prevention, Fake Cure, Fake Treatment, False Fact or Prevention, Correction/Calling out, Sarcasm/Satire, True Public Health, Response, False Public Health Response, Politics, Ambiguous/Difficult to Classify, Commercial Activity or Promotion, Emergency Response, News, Panic Buying
%    &
%    & \textbf{CHECK again}
\\\hline
% 8th
ReCOVery \cite{zhou2020recovery}
    & -Spread of fake news
        \newline -News, tweets
    & \scriptsize{\url{https://github.com/apurvamulay/ReCOVery}}
    & Distant Supervision
    & 2,029 news articles on coronavirus are finally collected in the repository along with 140,820 tweets, news content and social information revealing how it spreads on social media, which covers textual, visual, temporal, and network information
%    & \cite{weinzierl2021automatic, solovev2022moral, di2022health, shang2022duo}
%    & 87.3\% \cite{shang2022duo}
\\\hline
% 9th
Mega-COV \cite{abdul2020mega}
    & -Large diverse, longitudinal, and multi-lingual COVID-19
        \newline -Tweets
    & \scriptsize{\url{https://github.com/UBC-NLP/megacov}}
    & None
    & The dataset is diverse (covers 268 countries), longitudinal (goes as back as 2007), multilingual (comes in 100+ languages), and has a significant number of location-tagged tweets ($\approx$ 169M tweets)
%    & -
%    & -
\\\hline
% 10th
CoAID \cite{cui2020coaid}
    & -COVID-19 healthcare misinformation dataset
        \newline -News, media posts
    & \scriptsize{\url{https://github.com/cuilimeng/CoAID}}
    & None
    & It includes 5,216 news, 296,752 related user engagements, 958 social platform posts about COVID-19, and ground truth labels
%    & \cite{di2022health, pelrine2021surprising, guo2022survey}
%    & 99.26\% \cite{cui2020coaid}
\\\hline
% 11th
COVID-Fact \cite{liu2020adapting}
    & -Claims concerning the COVID-19 pandemic from Reddit
        \newline -Claims
    & \scriptsize{\url{https://github.com/asaakyan/covidfact}}
    & \revision{Manual (Crowd-sourced)}
    \newline Automatic filtering for true claims and refuted claims
    & The COVID-Fact dataset contains 4, 086 real-world claims with the corresponding evidence documents and evidence sentences to support or refute the claims. There are 1,296 supported claims and 2,790 automatically generated refuted claims
%    &
%    & 58.33\% \cite{liu2020adapting}
\\\hline
% 12th
FakeCovid \cite{shahi2020fakecovid}
    & -Multilingual cross-domain dataset for COVID-19
        \newline -News
    & \scriptsize{\url{https://github.com/Gautamshahi/FakeCovid}}
    & \revision{Manual (Authors)}
    & 5182 fact-checked news articles for COVID-19, collected from 04/01/2020 to 15/05/2020 Articles annotated into 11 different categories
%    & -
%    & 65\% \cite{shahi2020fakecovid}
\\\hline
% 13th
COVID19 Fake News Detection \cite{patwa2021fighting}
    & -Real and fake news on COVID-19
        \newline -Media posts, articles
    & \scriptsize{\url{https://github.com/parthpatwa/covid19-fake-news-detection}}
    & \revision{Manual (Authors)}
    & 10,700 social media posts and articles
%    & \cite{baris2021ecol}
%    & 98.13\% \cite{baris2021ecol}
\\\hline
% 14th
COVIDLIES \cite{hossain2020covidlies}
    & -COVID-19 related misinformation
        \newline -Tweets
    & \scriptsize{\url{https://github.com/ucinlp/covid19-data}}
    & \revision{Manual (Experts)}
    & 86 misconceptions and 6,761 annotated tweet-misconception pairs
%    & -
%    & 50.2\% \cite{hossain2020covidlies}
\\\hline
% 15th
CoVaxLies \cite{weinzierl2021automatic}
    & -COVID-19 vaccines
        \newline -Tweets
    & \scriptsize{\url{https://github.com/Supermaxman/covid19-vaccine-twitter}}
    & \revision{Manual (Experts)}
    & 7,246 tweets paired with Misinformation Targets (MisTs) and stances
%    & -
%    & 40.7\% \cite{weinzierl2021automatic}
\\\hline
% 16th
VaccineLies \cite{weinzierl2022vaccinelies}
    & -COVID-19 and HPV vaccines
        \newline -Tweets
    & \scriptsize{\url{https://github.com/Supermaxman/vaccine-lies}}
    & \revision{Manual (Experts)} \newline Automatic
    & 14,642 tweets paired with Misinformation Targets (MisTs) and stances (a superset of CoVaxLies)
%    & \cite{weinzierl2021automatic, weinzierl2022identifying}
%    & 87.1\% for stance classification \cite{weinzierl2022identifying}
\\\hline
% 17th
FakeHealth \cite{dai2020ginger}
    & -Real and fake health stories and release
        \newline -News, tweet engagements
    & \scriptsize{\url{https://github.com/EnyanDai/FakeHealth}}
    & \revision{Manual (Experts)}
    & 1,690 stories (1,218 real and 472 fake) and 606 releases (315 real and  291 fake)from HealthNewsReview.org, both news stories, and news releases are evaluated by experts on 10 criteria. Ratings from 0 to 5.
%    & \cite{di2022health}
%    & 80.2\% for releases and 75.6\% for stories\cite{dai2020ginger}
\\\hline
% 18th
COVID-19 Fake News Sentiment \cite{iwendi2022covid}
    & -COVID-19
        \newline -News
    & \scriptsize{\url{https://github.com/susanli2016/NLP-with-Python/blob/master/data/corona_fake.csv}}
    & \revision{Manual (Authors)}
    &Dataset consists of 586 true news and 578 fake news, and more than 1100 news articles and social media posts regarding COVID-19. The true news was gathered from Harvard Health Publishing, WHO, CDC, The New York Times, and so on. Fake news was gathered from social media (Facebook) posts and other medical sites.
%    & -
%    & 88\% \cite{iwendi2022covid}
\\\hline
% 19th
COVID-19 Disinformation \cite{alam2020fighting}
    & -COVID-19
        \newline -Tweets
    & \scriptsize{\url{https://github.com/firojalam/COVID-19-disinformation}}
    & \revision{Manual (Authors)}
    &Annotated dataset of 16K tweets related to the COVID-19 infodemic in four languages (Arabic, Bulgarian, Dutch, and English), using a schema that combines the perspective of journalists, fact-checkers, social media platforms, policymakers, and the society.
%    & \cite{kar2021no}
%    & 88\% \cite{kar2021no}
\\\hline
% 20th
MMCoVaR  \cite{chen2021mmcovar}
    & -COVID-19 vaccine
        \newline -News, Tweets
    & \scriptsize{\url{https://drive.google.com/drive/folders/1LGfD0mdup7va_SME0sncg-h3ADoCZyEZ?usp=sharing} (must request access)}
    & Automatic
    &Multimodal dataset, 2593 articles annotated as reliable or non-reliable and 24184 Twitter posts
%    & \cite{shang2022duo}
%    & 91.9\% on news and 85.8\% on tweets \cite{chen2021mmcovar}
\\\hline
% 21th
Covid-HeRA  \cite{dharawat2022drink}
    & -COVID-19
        \newline -Tweets
    & \scriptsize{\url{https://github.com/TIMAN-group/covid19_misinformation}}
    & Automatic for binary
        \newline Manual (Authors) for multiclass
    &62,286 tweets labeled as real news, possibly severe, highly severe, other, and refutes/rebuts (a subset of CoAID)
%    & -
%    & 33\% multiclass and 75\% for binary \cite{dharawat2020drink}
\\\hline
% 22th
CoVaxNet  \cite{jiang2022covaxnet}
    & -COVID-19 vaccine
        \newline -Tweets, Fact-checked Data, News (offline like local news), Statistics
    & \scriptsize{\url{https://github.com/jiangbohan/CoVaxNet}}
    & Manual \newline Automatic
    &A multi-source, multi-modal and multi-feature online and offline data repository. It contains 1,495,991 pro-vaccine and 335,229 anti-vaccine (labeled using keywords and hashtags), and 4,263 COVID-19 vaccine-related fact-checking reports
%    & -
%    & 79.3\% for claims \cite{jiang2022covaxnet}
\\\hline
% 23th
WICO  \cite{pogorelov2021wico}
    & -COVID-19 and 5G
        \newline -Tweets
    & \scriptsize{\url{https://datasets.simula.no/wico-graph}}
    & \revision{Manual (Authors)}
    &Subgraphs of 3,000 tweets that propagate 5G COVID-19 misinformation, other conspiracy theories, and tweets that do neither
%    & -
%    & 61.8\% for multiclass \cite{pogorelov2021wico}
\\\hline
% 24th
MM-COVID \cite{li2020mm}
    & -COVID-19
        \newline -News
    & \scriptsize{\url{https://github.com/bigheiniu/X-COVID}}
    & Manual
    &3,981 fake and 7,192 trustworthy multi-language and multi-modal news, including temporal information
%    & -
%    & 90\% \cite{li2020mm}
\\\hline
\end{longtable}
\end{footnotesize}

\subsection{Fact-Checking Tools}
\label{sec:tools}
Table \ref{tab5} lists the most popular fact-checking tools. For every tool, it reports whether it uses manual or automatic fact-checking techniques, the themes they are covering, and the rating scale they use for the credibility assessment of the news. Below we discuss the most prominent ones.
%Table \ref{tab6} reports their traffic statistics.

Snopes \cite{Snopes} is one of the oldest fact-checking websites which has been used as a source for validating and debunking urban legends and stories in American culture. It covers politics, health, and social issues. In 2020 Snopes was very active regarding the COVID-19 pandemic misinformation and had around 237 COVID-related fact-checked articles. Its rating scale consists of  values like \emph{True, Mostly True, Mixture, Mostly False, Unproven}, etc.

Media Bias/Fact Check \cite{MediaBias} examines bias in media from all points of the political spectrum. It includes a "Daily Source Bias Check" that examines the truthfulness and bias of various news sources. It also includes a Conspiracy / PseudoScience category that includes COVID-19-related news.

Another well-known fact checker is Politifact \cite{Politifact}. This service is devoted to fact-checking claims made by political pundits. Specifically, everyday journalists explore transcripts, speeches, news stories, press releases, and campaign brochures, to identify statements that should be fact-checked. In order to decide which statements to check they consider questions like "Is the statement rooted in a fact that is verifiable?" or "Is the statement likely to be passed on and repeated by others?". They cover politics and health topics, and they provide a scorecard that gives information about the statistics on the authenticity distribution of all statements related to a specific topic.

The Washington Post \cite{washington-post-fact-checker} also provides a column discussing the factual accuracy of statements with zero to four pinocchios. Full-fact \cite{full-fact} is another fact checker that uses a three-stage review fact-checking process that may also involve external academics (soon it will also apply automatic approaches). FactCheck \cite{fact-check} consists of features like “Ask FactCheck” with which users can ask questions that are usually based on rumors, “Viral Spiral” which makes a list of the most popular myths that the site has debunked, “Mailbag” which includes readers’ opinion on the site claims.

Finally, Google Fact Check \cite{google-fct} is a search engine for fact checks. It consists of the Fact Check Markup tool which provides structured markup and the Fact Check Explorer. The first one allows publishers to add claim-related markup into their web content which will then be crawled and indexed by Google. The explorer allows users to query claims and returns whether they are true or false based on the indexed structured markup added by publishers and Google’s algorithms that evaluate the credibility of the sources.

\begin{footnotesize}
\begin{table*}[t]
\caption{Fact-checking tools}
\label{tab5}
\begin{tabular}{|l|l|l|p{150pt}|}
\hline
\textbf{Name}
    &  \textbf{Manual / Automatic}
    & \textbf{API}
    & \textbf{Themes}
    %& \textbf{Rate scale}
\\\hline\hline
% 1st
Snopes
    & Manual
    & -
    & Politics, Health, social issues
    %& True, Mostly True, Mixture, Mostly False, False, Unproven, Unfounded, Outdated, Miscaptioned, Correct attribution, Legend, Misattributed, Scam, Legit, Labeled Satire, Originated as Satire, Recall, Lost Legend
\\\hline
% 2nd
Media Bias/Fact Check
    & Manual
    & -
    & Science, Politics
    %& Least Biased, Left Bias, Left-Center Bias, Right Center Bias, Right Bias, Conspiracy-Pseudoscience, Questionable Sources, Pro-Science, Satire
\\\hline
% 3rd
PolitiFact
    & Manual
    & \checkmark
    & Politics, Health
    %& True, Mostly true, Mostly false, False
\\\hline
% 4th
The Washington Post
    & Manual
    & -
    & Politics, Technology, Lifestyle, Word, Sports, Health, Well Being, Science
    %& One Pinocchio, Two Pinocchio, Three Pinocchio, Four Pinocchio, The Geppetto checkmark, An upside-down Pinocchio, Verdict pending
\\\hline
% 5th
Full Fact
    & Manual \& soon Automatic
    & \checkmark
    & Health, Economy, Crime, Law, Education, Europe
    %& -
\\\hline
% 6th
FactCheck
    & Manual
    & -
    & Politics, Science, Health
    %& True, No evidence, False
\\\hline
% 7th
LeadStories
    & Manual
    & -
    & Politics, Health
    %& No clear labels
\\\hline
% 8th
ClaimBuster
    & Automatic
    & \checkmark
    & General
    %& -
\\\hline
% 9th
Emergent
    & Manual
    & -
    & Culture, Business, Politics, Viral
    %& True, False, Unverified
\\\hline
% 10th
TruthOrFiction
    & Manual
    & -
    & Politics, Health, Nature
    %& Truth, Fiction, Unproven, Truth \& Fiction, Pending investigation
\\\hline
% 11th
Google Fact Check
    & Manual \& Automatic
    & \checkmark
    & General
    %& True/False or description of assessment
\\\hline
\end{tabular}
\end{table*}
\end{footnotesize}

%%%%%%%%%%%%%%%%%%%%%%%%%%%%%%%%%%%%%%%%%%%%%%%%%%%%

% COMMENTED_OUT
\iffalse %%%%%%%%%%%%%%%%%%%%%%%%%%%%%%%%%%%%%%%%%%%%
\subsection{Summary}
In the following, Table \ref{tab2} reports the reviewed scientific papers related to the detection of medical fake news, their context, sources of data, and the themes they are covering, along with the approaches and features they have used. Subsequently, Table \ref{tab3} reports more details about the characteristics of the datasets used in those papers, like their attributes and their sources.
\fi%%%%%%%%%%%%%%%%%%%%%%%%%%%%%%%%%%%%%%%%%%%%%%%%%

\section{Mitigation}
\label{sec:mitigation}
Fake news and misinformation shared on social media can be very effective in shaping opinions and beliefs of people. Medical fake news, in particular, is a serious danger to public health, as the spread of false information and lack of knowledge of the truth can put patients at risk. The proliferation of medical fake news can cause individuals to adopt dangerous health practices or avoid proven, life-saving medical treatments.
%, which can lead to serious consequences.
As such, it is imperative that the spread of medical fake news on social media is effectively mitigated using robust solutions that can limit its impact on public health. The issue of mitigating fake news has garnered attention from researchers across various fields, including social sciences and psychology. However, in this survey, we will only describe the technical approaches developed in the field of computer science.

One category of mitigation methods revolves around the user perspective, leveraging the diverse roles and positions users can have in the dissemination of fake news. Some users hold significant influence as opinion leaders and have a wide circle of followers, while others act as correctors and contribute to fake news posts by attaching links and comments that counter the false information with facts and evidence \cite{zhou2020survey}.

%tsap: Isnt the content in this paragraph included in the source-based methods?
Other approaches involve the identification of sources that create and disseminate disinformation, which are afterward prevented from contaminating social media platforms as described in  Section~\ref{sec:sources}. Such approaches are characterized by their proactive nature and can be included in the early fake news detection methods, as they aim to identify users who are at risk of engaging with fake news or have previously engaged with it and may do so again in the future. These techniques utilize content-based methods, like the ones described in \ref{sec:content} to analyze the language and content patterns of news articles in the user history and their personal information. Through the application of these techniques, it becomes feasible to identify and remove fake users or restrict their movements on social media platforms. An important aspect of these techniques is the identification of influential spreaders. When blocking fake news in a social network, targeting these influential spreaders can lead to a more efficient intervention compared to those with insignificant social influence on others \cite{saxena2022fake}.

Bot detection techniques described in Section~\ref{sec:sources} also belong to this category. These methods focus on identifying and removing fake users that are designed to mimic human behavior and disseminate misinformation by detecting patterns of bot behavior, such as repeated posting of similar content or engagement with other bots. Once detected, these fake users can be removed from social media platforms, limiting their ability to spread false information \cite{10.1145/2818717,egli2021bad,heidari2021bert}.

Also, \cite{vo2018rise} presents an innovative approach to effectively mitigate the problem of online misinformation. This study delves into the role of ``guardian'', online users who actively combat misinformation and fake news during online discussions by providing fact-checking URLs. The authors also introduce a novel fact-checking URL recommendation model, designed to motivate and empower guardians to engage even more actively in fact-checking activities.

Intervening based on network structure involves stopping the spread of fake news by blocking its propagation paths \cite{saxena2022fake}. This approach relies on analyzing the network structure of how fake news spreads and predicting how it will continue to spread. By identifying the key nodes in the network through which the fake news is likely to spread, these nodes can be blocked in order to prevent the spread of misinformation.
These approaches employ diverse models, such as the Independent Cascade (IC) \cite{wu2017scalable, kempe2005influential,saxena2015understanding} and Linear Threshold (LT) \cite{wu2016mining,yang2020containment} models, to model the diffusion of information in the network. Also, network structural and temporal features are used in order to detect the propagation of fake news and slow it down.

Another approach to limit the propagation of fake news on social media aims to identify the smallest possible set of users in the network whose immunization can effectively diminish the spread of misinformation \cite{pham2018targeted}. To find these nodes, various methods and techniques are utilized in graphs. One such technique is the greedy method, which proceeds iteratively, selecting each time the best node that minimizes the network influence, until the desired number of nodes is reached. Other techniques rely on heuristic methods. Additionally, measures such as centrality measures can be employed to estimate the magnitude of influence that a node has on other nodes in the network, which can help in identifying nodes with the highest influence to be immunized.

Detecting and monitoring communities in social networks is also crucial in identifying the ones where fake news is frequently shared and has the potential to spread beyond the community. In \cite{wu2019minimizing} researchers developed a method to measure the influence of nodes within a community and calculate the probability of rumors spreading between them. They also suggested an algorithm that employs a dynamic blocking strategy to minimize the impact of rumors.

Some other network approaches focus on mitigating fake news by promoting real news, by either targeting users who have already been exposed to false news, or by strategically selecting initial users to disseminate true news in competition with false news.
%These methods aim to counteract the spread of false news within a network by promoting accurate information.
An example is the decontamination method described in \cite{nguyen2012containment}, which targets contaminated users (i.e., users who have been exposed to false news) for decontamination.
The contaminated users are identified using greedy algorithms while the IC and LC models are used for the diffusion process. The process of selecting seed users continues until the expected objective of decontaminating a $\beta$-fraction of the contaminated users is reached. Another approach named the Competing Cascades method \cite{saxena2022fake} involves selecting $k$ seed users who will propagate the true news cascade. The objective is to maximize the influence of the true news cascade while minimizing the influence of the fake news cascade.
%The objective of this strategy is to minimize the number of users who are influenced by the fake news article and maximize the number of users who are influenced by the true news article.
The users who are exposed to both articles may be more likely to believe the true news article if it is shared by a trusted source or if it provides compelling evidence.

\section{Discussion and Future Work}
\label{sec:discuss}
In the recent years,  research advances in addressing misinformation, and specifically health misinformation,  were expedited due to the COVID pandemic. However there are still several open issues and major challenges in the domain of medical misinformation including the following areas \cite{suarez2021prevalence}:
%The authors in  \cite{suarez2021prevalence} pinpoint the following major challenges in the domain of medical misinformation:
(a) the identification of dominant health misinformation trends and their prevalence across different social platforms, (b) the mechanisms that allow the spread of health misinformation, (c) their impact on the development and reproduction of unhealthy and dangerous behaviors, and finally, (d) the development of strategies for fighting and reducing the negative impact of misinformation without reducing the inherent communicative potential to propagate health information. Addressing the various open problems in these areas  requires the collaboration of \textit{cross-discipline} researchers, policymakers, and technology experts.

\revision{One challenge specific to the medical domain is the \textit{inherent uncertainty} of medical scientific knowledge and the fact that scientific knowledge \textit{evolves over time} \cite{medical-uncertainty}. Medical facts are usually accompanied with probabilities and risk factors that may differ based on population demographics.
Furthermore, medical knowledge evolves over time, as new discoveries are made. This uncertainty poses additional challenges in communicating medical information to the public, and provides a fertile ground for misinformation to grow.
Uncertainty and temporality
should be accounted for when designing approaches for fact-checking, and misinformation detection or mitigation.}

Another challenge is the fact that the approaches used for  fake news fabrication and propagation in social media and networks are constantly improving. Aiming at accurately imitating the characteristics and features of valid news as closely as possible, the deployed strategies are becoming more complex and refined. The recent advancements in large language models (LLMs), which can be trained to follow the style and form of valid news  will only complicate the problem \cite{LLM}.
Consequently, the assumption that fake news has distinct characteristics that separate it from valid news weakens, as producers and promoters of fake news deploy more advanced techniques.
%The same holds for propagation-based methods where fake news-related patterns can be adapted to imitate the structural, temporal, and other features of valid news propagation networks.
\revision{There are similar challenges for propagation-based methods where bots can  be adapted to imitate the structural, temporal, and propagation patterns of valid news propagation networks. However, faithfully reproducing human behavior is more complex, so these features can still preserve some of their discriminatory power.}

%Toward this end,
%we anticipate that future work in fake news detection will mainly focus on exploiting and incorporating
An important future research direction for addressing this challenge is to incorporate available authoritative ground truth knowledge and apply symbolic logic in fake news detection. Specifically for the medical domain, hybrid approaches are needed that combine the constantly evolving machine mearning approaches for detection, with the slower-evolving, but high-quality and validated information stored in domain-specific Knowledge Bases. Such information usually originates from authoritative academic research and can be exploited by logic rules and reasoning. The fusion of the statistical learning capabilities of the deep neural networks with strict logic reasoning approaches and rules (e.g., \cite{wang2020integrating}), will lead to a representation learning system with a joint learning mechanism, that will achieve joint inference and will enforce complicated correlations and constraints in the output space.

Regarding the focus of research in medical fake news, most efforts  focus
on fake news detection.
%which despite being a crucial task does not deter the propagation of fake news.
Detection needs to be combined with mitigation approaches, which act as gatekeepers of misinformation propagation.
%We expect further
More research effort is required for the development and refinement of mitigation approaches, along with their evaluation in real-life settings.
However, we advocate that there is a pressing need for mitigation approaches to also provide transparency, explainability, and fairness guarantees, to build a relationship of trust between the producers, the consumers, and the gatekeepers of the information flow (i.e., the social media platforms). The goal should be the creation of a healthy ecosystem around grounded information that does not hinder freedom of speech, diversity, and exploration.

In the same direction, the research community should work towards preemptive approaches that provide valid and fact-checked information, personalized to the current interests of users. The goal is to proactively inform a user about the facts of a misinformation-infested topic, which could potentially attract the attention of the user in the near future. This way, the user will have a critical and well-informed view, if and when exposed to this misinformation. Furthermore, inoculated users will act as deterrent to the propagation and spread of misinformation in the social network.

 Finally, an important finding of this work is that there is not a single specific dataset in the domain of medical fake news that can be used as a benchmark.  Most of the described approaches introduce their own customized datasets, making the comparison of the proposed methods difficult. Moreover, the quality of the gathered data and their annotation varies and is sometimes questionable. The research community should focus on collaboratively creating a reference dataset for the task of fake news detection in the medical domain. This dataset should incorporate rich data gathered from social media and the Web and should satisfy the requirements of content-based, propagation-based, and source-based misinformation detection methods. The dataset should cover a broad range of medical topics, beyond COVID-19,  such as smoking, drugs, diets, and self-treatment-related content. Finally, the important work of annotating the dataset with labels should be done by experts, incorporating data from authoritative expert-based manual fact-checking services.

\section{Conclusion}
\label{sec:conc}
Understanding, assessing, and countering health misinformation promises to empower healthcare providers and policymakers, safeguard the public trust in medical information science, enhance and promote health literacy, and ultimately contribute to improved health outcomes. While ongoing collaboration between researchers, policymakers, and healthcare providers is essential, technology offers the means to more objectively and reliably address misinformation.

In this comprehensive survey we shed light on the pervasive issue of health misinformation online, with an emphasis on the COVID-19 pandemic.
We provide an overview of the domain and examine notable instances of medical fake news and its consequences, highlighting the importance of addressing this critical challenge.
We also provide an extensive analysis of manual and automatic methods for fact-checking, along with content, propagation, and source-based techniques for fake news detection, and automated methods for misinformation mitigation.
%it becomes evident that
Moreover, we present existing publicly available datasets specifically tailored to fact-checking and fake news detection in the medical domain, and we detail a collection of well-known fact-checking tools. These resources serve as a foundation for further research and development in the field, enabling researchers and practitioners to refine and enhance the effectiveness of their approaches.

%%%%%%%%%%%%%%%%%%%%%%%%%%%%%%%%%%%%%%%%%

While significant progress has been made, it is crucial to acknowledge that the battle against health misinformation is an ongoing endeavor.
%Continued collaboration between researchers, policymakers, and technology experts is essential to stay ahead of the evolving landscape of misinformation.
Future directions include the integration of machine learning and natural language advancements, the more extensive deployment and exploitation of knowledge graphs, the exploration of novel detection techniques, the development of comprehensive curated datasets that can act as benchmarks, and the further development and refinement of mitigation and preemptive approaches.

%By addressing the challenge of fake news in the medical domain effectively, we can safeguard the public trust in medical information, enhance health literacy, and ultimately contribute to a more informed and resilient society.

\begin{acks}
This research was conducted as a collaboration between the University of Ioannina and Pfizer. Pfizer is the research sponsor.
\end{acks}


\bibliographystyle{ACM-Reference-Format}
\bibliography{ref}

\end{document}